\documentclass[
aps,
prc,
superscriptaddress,
floatfix,
nofootinbib,
twocolumn
]{revtex4-2}

\usepackage{amsmath,amssymb,amsfonts,physics,graphicx,float}
\usepackage[setpagesize=false]{hyperref}
\allowdisplaybreaks[4]

\usepackage{subcaption} 

\usepackage{ulem}
\usepackage{color}
\definecolor{ar}{rgb}{1.0, 0.01, 0.24}
\definecolor{al}{rgb}{0.82, 0.1, 0.26}
\definecolor{ev}{rgb}{0.56, 0.0, 1.0}

\begin{document}

\title{
Gravitational wave spectrum
from first-order QCD phase transitions based on a parity doublet model
}

\author{Bikai Gao}
\email{bikai@rcnp.osaka-u.ac.jp}
\affiliation{Research Center for Nuclear Physics, Osaka University, Ibaraki, Osaka 567-0047, Japan}

\author{Jingdong Shao}
\email{shaojingdong19@mails.ucas.ac.cn}
\affiliation{
School of Physical Sciences, University of Chinese Academy of Sciences, Beijing 100049, China}

\author{Hong Mao}
\email{mao@hznu.edu.cn}
\affiliation{School of Physics, Hangzhou Normal University, Hangzhou 311121, China}


\date{\today}

\begin{abstract}
We investigate the gravitational wave spectrum from first-order QCD phase transitions using the parity doublet model at finite baryon chemical potential. The model incorporates the chiral invariant mass $m_0$, representing the portion of nucleon mass that persists even when chiral symmetry is restored. Within the model, we identify two first-order phase transition regions: the nuclear liquid--gas transition and the chiral phase transition. By solving the bounce equation and computing the Euclidean action $S_3/T$, we obtain the gravitational wave spectra from both transitions. The liquid--gas transition yields $\alpha \sim \mathcal{O}(1)$ and $\beta/H \sim \mathcal{O}(10)$--$\mathcal{O}(100)$ near the endpoint of the first-order line, producing signals with peak frequencies from the millihertz to the nanohertz band that can fit the existing data. In contrast, the chiral transition produces signals suppressed by approximately five orders of magnitude, well below the sensitivity of all current and planned detectors. These results connect the chiral invariant mass to the gravitational wave spectrum, offering a novel probe of the origin of nucleon mass through gravitational wave astronomy.
\end{abstract}

\maketitle



\section{Introduction}
\label{sec:intro}

The quantum chromodynamics (QCD) phase transition plays a crucial role in understanding both the evolution of the early Universe and the internal structure of compact stars. 
At high temperatures and/or densities, QCD matter is expected to undergo phase transitions associated with chiral symmetry restoration and color deconfinement. 
While lattice QCD simulations have established that the transition at vanishing or small baryon chemical potential is a smooth crossover, 
the nature of the QCD phase transition at finite baryon density remains an open question due to the sign problem that plagues lattice calculations in this regime.

First-order phase transitions in the early Universe are of particular interest as they can generate stochastic gravitational wave (GW) backgrounds through several mechanisms: 
bubble nucleation and collision, sound waves in the plasma, and magnetohydrodynamic turbulence. 
The recent observations of a stochastic GW background in the nanohertz frequency band by multiple pulsar timing array (PTA) collaborations, 
including NANOGrav~\cite{NANOGrav:2023hde,Lerambert-Potin:2021ohy}, EPTA/CPTA~\cite{Zic:2023gta,Reardon:2023gzh,EPTA:2023sfo,EPTA:2023fyk,Xu:2023wog} and PPTA~\cite{Xue:2021gyq},
have stimulated significant interest in exploring cosmological sources of such signals. 
Among the proposed explanations, first-order QCD phase transitions represent a compelling possibility, 
though the predicted GW spectra are highly sensitive to the phase transition dynamics, particularly the transition rate $\beta/H$.

In standard cosmological scenarios, the cosmic QCD phase transition from the quark-gluon plasma to the hadronic phase occurred approximately $10$ microseconds after the Big Bang at temperatures around $T_c \sim 150$--$170$ MeV. 
For physical quark masses, this transition is expected to be a crossover. 
However, first-order phase transitions can occur in several scenarios: 
the chiral phase transition in the massless three-flavor limit~\cite{Pisarski:1983ms}, 
the deconfinement transition in pure gauge theory~\cite{Svetitsky:1982gs}, 
systems with chirality imbalance~\cite{Yu:2014sla,Shao:2023gho}, 
and importantly, at high baryon chemical potentials where a critical point may separate the crossover from a first-order transition region~\cite{Fukushima:2010bq,Stephanov:2004wx}.

The generation of high baryon density in the early Universe, while seemingly at odds with the observed small baryon asymmetry $\eta_B = n_B/s \sim 10^{-9}$, 
can be naturally achieved through mechanisms such as Affleck-Dine baryogenesis~\cite{Affleck:1984fy,Dine:2003ax}. 
The initially large baryon asymmetry can be subsequently diluted by  little inflation~\cite{Borghini:2000yp,Boeckel:2009ej,Boeckel:2010bey,Schettler:2010dp,Boeckel:2011yj} to match observations, 
allowing for first-order QCD phase transitions to occur in the early Universe. 
This scenario opens the possibility for generating detectable GW signals from QCD phase transitions.

Recent studies have investigated GW spectra from first-order QCD phase transitions using various effective models, 
including the Friedberg-Lee model~\cite{PhysRevD.18.2623,PhysRevD.16.1096,PhysRevD.15.1694}, 
and the quark-meson (QM) and Polyakov-quark-meson (PQM) models~\cite{Wang:2023omt,Wang:2023pmn,Gupta:2011ez,Shao:2024dxt}. 
These effective QCD theories have revealed that the transition rate $\beta/H$ is typically large ($\sim 10^4$--$10^5$) at high temperatures with small chemical potential~\cite{Chen:2022cgj,Morgante:2022zvc,Helmboldt:2019pan}, 
yielding GW peak frequencies in the $10^{-4}$--$10^{-2}$ Hz range detectable by future space-based interferometers such as LISA and Taiji. 
At high baryon chemical potentials, the transition rate can be significantly reduced, potentially producing nanohertz GWs observable by PTAs.

In this work, we employ the parity doublet model to investigate the GW spectrum from first-order chiral phase transitions. 
The parity doublet model~\cite{Detar:1988kn,Jido:2001nt,Gallas:2009qp,Steinheimer:2011ea} provides a unique framework for describing baryonic matter 
that naturally incorporates the concept of chiral symmetry restoration while maintaining a nonzero nucleon mass even in the chirally restored phase. 
In this model, the nucleon and its parity partner $N(1535)$ form a chiral doublet, and the nucleon mass receives contributions from two distinct sources: 
the chiral invariant mass $m_0$, which persists even when chiral symmetry is restored, 
and the contribution from spontaneous chiral symmetry breaking proportional to the chiral condensate $\langle\bar{q}q\rangle$~\cite{Zschiesche:2006zj,Dexheimer:2007tn,Sasaki:2011ff,Steinheimer:2011ea,Motohiro:2015taa,Gao:2025eax}. The parity doublet model offers several advantages for studying QCD phase transitions and their astrophysical implications. 
The model has been successfully applied to describe neutron star properties, 
and the chiral invariant mass $m_0$ can be constrained by neutron star observations from NICER and GW detections from LIGO/Virgo~\cite{Marczenko:2017huu,Minamikawa:2020jfj,Marczenko:2021uaj,Gao:2024chh,Gao:2025nkg,Gao:2025vdc,Gao:2025okn,Yuan:2025dft,Fraga:2023wtd,Gao:2026scv}. 
Also, the model naturally exhibits a first-order chiral phase transition at finite baryon density for a wide range of parameters, 
making it particularly suitable for studying bubble nucleation and GW generation. 
Furthermore, the chiral invariant mass $m_0$ provides a direct connection to the fundamental question of the origin of hadron mass, 
as it represents the portion of nucleon mass that does not arise from spontaneous chiral symmetry breaking.

This paper is organized as follows. In Sec.~\ref{sec-1}, we briefly introduce the parity doublet model and present the corresponding phase diagram. In Sec.~\ref{sec_3}, we discuss the dynamics of homogeneous nucleation. In Sec.~\ref{sec_4}, we compute the resulting GW spectrum. Finally, in Sec.~\ref{sec-summary}, we summarize our results and discuss their implications and possible future directions.


\section{Parity doublet model}\label{sec-1}
In this section, we briefly review the construction of parity doublet model. The thermodynamic potential at finite temperature $T$ is given as~\cite{Motohiro:2015taa,Marczenko:2025kpv}
\begin{align}
\Omega = \Omega_H + V(\sigma) - V_0(\sigma) + V_{\omega},
\end{align}
with
\begin{equation}
\begin{aligned}
    \Omega_H&=- 2T\sum_{\alpha=\pm}\sum_{i=p,n}\int \frac{d^3 p}{(2\pi)^3}\left[\ln(1 - f_{\alpha, i}) + \ln(1 - \bar{f}_{\alpha, i}) \right], \\
    V(\sigma) &=-\frac{1}{2}\bar{\mu}^2 \sigma^2 + \frac{1}{4}\lambda \sigma^4 - \frac{1}{6}\lambda_6 \sigma^6 - m_\pi^2f_\pi \sigma,\\
    V_0(\sigma) &=-\frac{1}{2}\bar{\mu}^2 f_\pi^2 + \frac{1}{4}\lambda f_\pi^4 - \frac{1}{6}\lambda_6 f_\pi^6 - m_\pi^2f_\pi^2,\\
    V_\omega &= -\frac{1}{2}m_\omega^2\omega^2.
\end{aligned}
\end{equation}
Here $\alpha=\pm$ correspond to positive- and negative-parity nucleon state and $f_{\alpha, i} ( \bar{f}_{\alpha,i} )$ denote for Fermi-Dirac distribution function expressed as
\begin{align}
f_{\pm,i} &= \frac{1}{1 + e^{ (E_\pm - \mu_B^*)/T}},\\
\bar{f}_{\pm, i} &=\frac{1}{1 + e^{ (E_\pm + \mu_B^*)/T}} 
\end{align}
where $T$ is the temperature, the single-particle energy $E_\pm = \sqrt{{\bf p}^2 +m_\pm^2}$ and the effective baryon chemical potential is defined as $\mu_B^* = \mu_B - g_{\omega NN} \omega$. The effective mass $m_\pm$ takes the form
\begin{align}
    m_\pm = \frac{1}{2}\left[\sqrt{(g_1 + g_2)^2\sigma^2 + 4m_0^2} \pm (g_1 - g_2) \sigma \right]
\end{align}
with $m_0$ the chiral invariant mass and $g_1, g_2$ the Yukawa coupling constants determined from the vacuum mass of $N(939)$ and $N(1535)$. The mean fields $\sigma, \omega$ here are determined by following stationary conditions
\begin{align}
    \frac{\partial \Omega}{\partial \sigma} = \frac{\partial \Omega}{\partial \omega}=0,
\end{align}
with the rest model parameters are determined by fitting to zero temperature nuclear saturation data.
Tab.~\ref{input: mass} lists the hadron masses and pion decay constant used as inputs, while Tab.~\ref{saturation} summarizes the nuclear matter saturation properties at $n_0 = 0.16$ fm$^{-3}$. For each $m_0$ choices, the determined parameters are exactly the same as listed in Ref.~\cite{Minamikawa:2020jfj}.
\begin{table}[htbp]
\centering
	\caption{  {\small Physical inputs in vacuum in unit of MeV.  }  }\label{input: mass}
	\begin{tabular}{cccccccc}
		\hline\hline
		~$m_\pi$~&~ $f_\pi$ ~&~$m_\eta$ ~&~ $m_{a0}$ ~&~ $m_\omega$ ~&~ $m_\rho$ ~&~ $m_+$ ~&~ $m_-$\\
		\hline
		~140 ~&~ 92.4 ~&~ 550 ~&~ 980 ~&~ 783 ~&~ 776 ~&~ 939 ~&~ 1535\\
		\hline\hline
	\end{tabular}
\end{table}	
\begin{table}[htbp]
\centering
	\caption{  {\small Saturation properties used to determine the model parameters: the saturation density $n_0$, the binding energy $E_{\rm Bind}$, the incompressibility $K_0$, symmetry energy $S_0$. 
 }  }
	\begin{tabular}{cccc}\hline\hline
	~$n_0$ [fm$^{-3}$] ~& $E_{\rm Bind}$ [MeV] ~& $K_0$ [MeV] ~& $S_0$ [MeV] ~\\
	\hline
	0.16 & 16 & 240 & 31 \\
	\hline\hline
	\end{tabular}
	\label{saturation}
\end{table}	
From the thermodynamic relations, we obtain the pressure and the corresponding energy density as
\begin{equation}
\begin{aligned}
    p &= -\Omega, \\
    \varepsilon & = -p + T \frac{\partial p}{\partial T} + \mu_B n_B. 
\end{aligned}
\end{equation}

\begin{figure}[htbp]
\centering
\includegraphics[width=1\hsize]{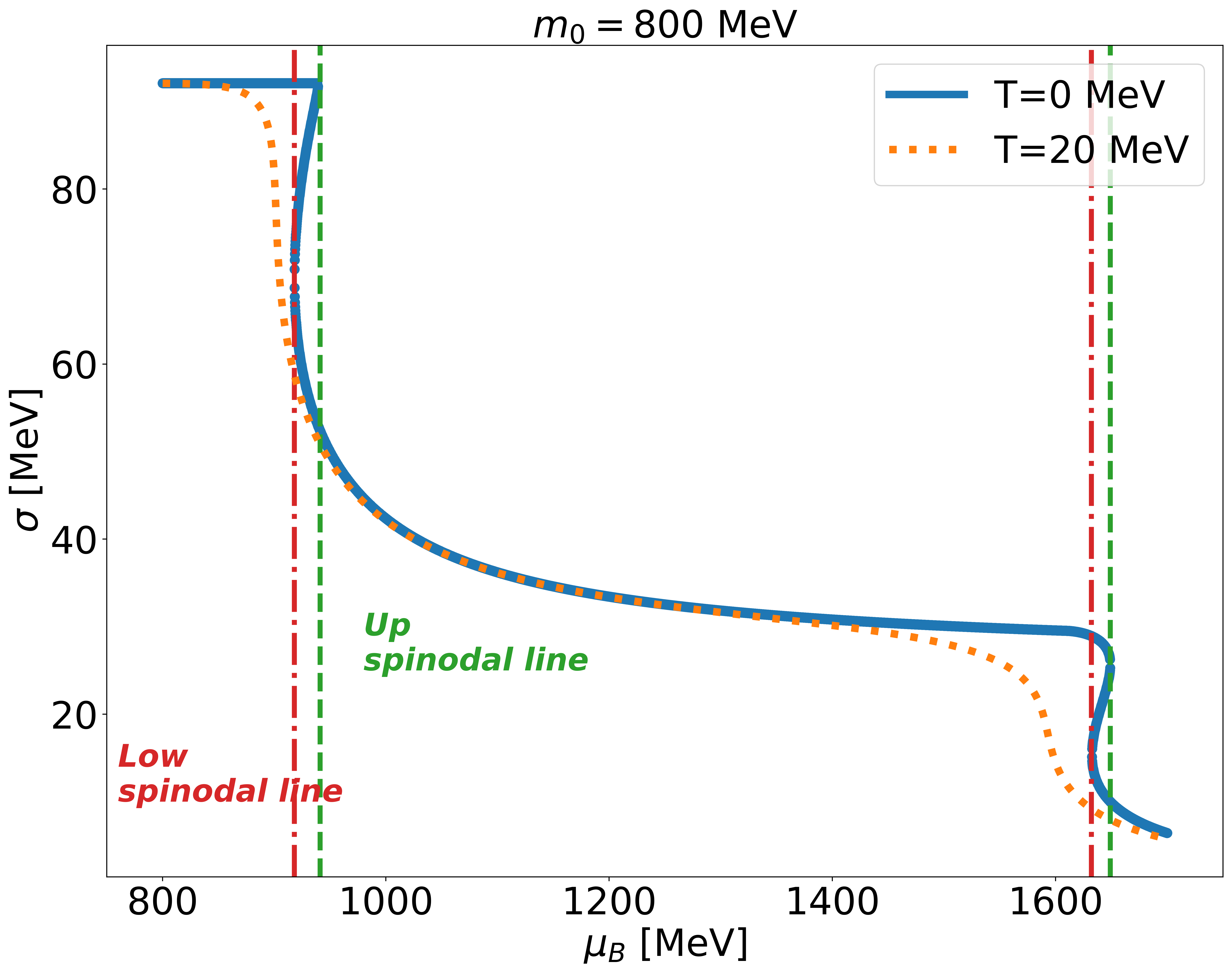}
\caption{Expectation value of the $\sigma$ mean field as a function of baryon chemical potential for $T = 0$ and $20$ MeV.}
\label{si_mub}
\end{figure}

In this study, we fix the $m_0=800$ MeV which is consistent with the constraints from the recent neutron star observations ~\cite{Minamikawa:2020jfj,Gao:2024chh,Gao:2025nkg,Kong:2025dwl}.
Figure~\ref{si_mub} show the expectation value of the $\sigma$ mean field as a function of baryon chemical potential $\mu_B$ at two representative temperatures: $T = 0$ MeV (blue curve) and $T = 20$ MeV (orange curve). Two distinct phase transitions are observed. The first, occurring in the low-density region, corresponds to the nuclear liquid--gas phase transition, while the second, at higher chemical potential, is the chiral phase transition associated with the entering of the excited nucleon state $N(1535)$.
At zero temperature, both transitions are of first order, as evidenced by the discontinuous jumps in the $\sigma$ field. The upper and lower spinodal lines, indicated in the figure, define the metastable region where the gap equation for $\sigma$ admits multiple solutions at fixed $\mu_B$. These solutions correspond to the true vacuum (TV) and the false vacuum (FV) of the thermodynamic potential---a structure essential for bubble nucleation during cosmological phase transitions. As the temperature increases, the discontinuities become less pronounced, and the first-order transitions eventually turn into smooth crossovers, as illustrated by the orange curve.

\begin{figure}[htbp]
\centering
\includegraphics[width=1\hsize]{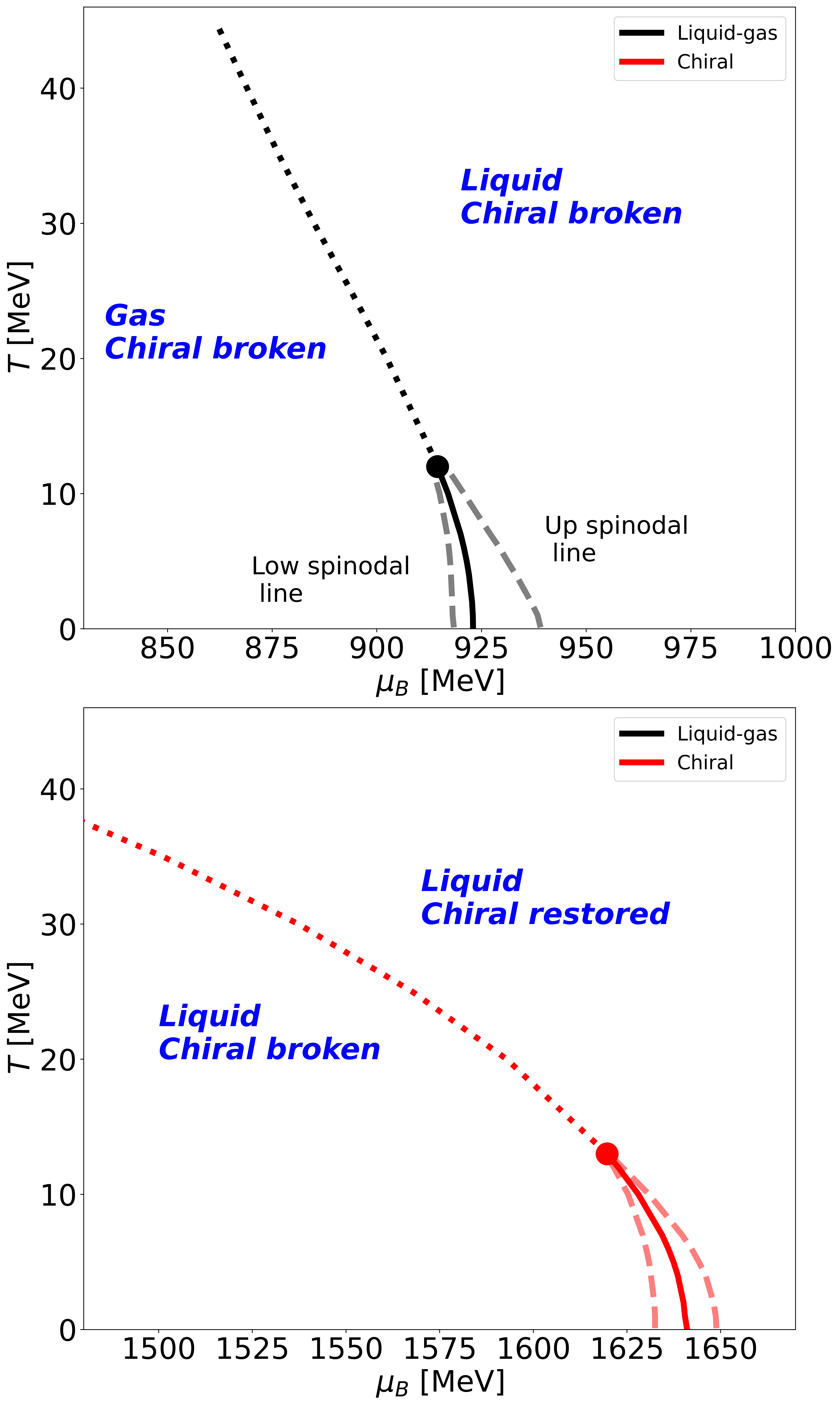}
\caption{Low-temperature phase diagram from the parity doublet model. The upper panel shows the region near the liquid-gas phase transition, while the lower panel presents the region near the chiral phase transition.}
\label{phase_dia}
\end{figure}

The low-temperature phase diagram derived from the parity doublet model is presented in Fig.~\ref{phase_dia}, with the upper panel showing the region near the liquid--gas phase transition and the lower panel displaying the region near the chiral phase transition. For each transition, the spinodal lines denoted by the dashed curves enclose a coexistence region where two local minima of the thermodynamic potential exist. This metastable region, bounded by the spinodal lines, is precisely where bubble nucleation can occur during a first-order phase transition.

\section{HOMOGENEOUS THERMAL NUCLEATION}\label{sec_3}
Cosmic first-order phase transitions are driven by the formation of bubbles containing the true vacuum. When the Universe cools down, bubbles of the lower-energy true vacuum nucleate within the surrounding false vacuum. The pressure difference between these two states drives the rapid expansion of the bubbles, transferring released energy into the bubble walls. When these expanding bubbles eventually collide, they dissipate kinetic energy into the cosmic medium. This generates GWs from three primary sources: the physical collision of bubble walls, sound waves in the plasma, and magnetohydrodynamic (MHD) turbulence.

The dynamics of bubble nucleation can be described within the framework of homogeneous thermal nucleation theory. The nucleation rate per unit volume is given by~\cite{Coleman:1977py,Ellis:2020awk,Binetruy:2012ze}
\begin{align}
    \Gamma(t) = A(t) e^{-S_4(t)}.
\end{align}
where $S_4$ is the four-dimensional Euclidean action evaluated on the $O(4)$-symmetric bounce solution, and $A(t)$ is the prefactor. At high temperatures ($T \gg R_c^{-1}$, where $R_c$ is the critical bubble radius), the bounce solution becomes approximately time-independent, and the action reduces to the three-dimensional Euclidean action $S_3/T$ evaluated on the $O(3)$-symmetric bounce solution. In this limit, the prefactor takes the form~\cite{Eichhorn:2020upj}
\begin{align}
    A(T) = T^4 \left(\frac{S_3}{2\pi T}\right)^{\frac{3}{2}}.
\end{align}

\begin{figure*}[htbp]
\centering
\includegraphics[width=1\hsize]{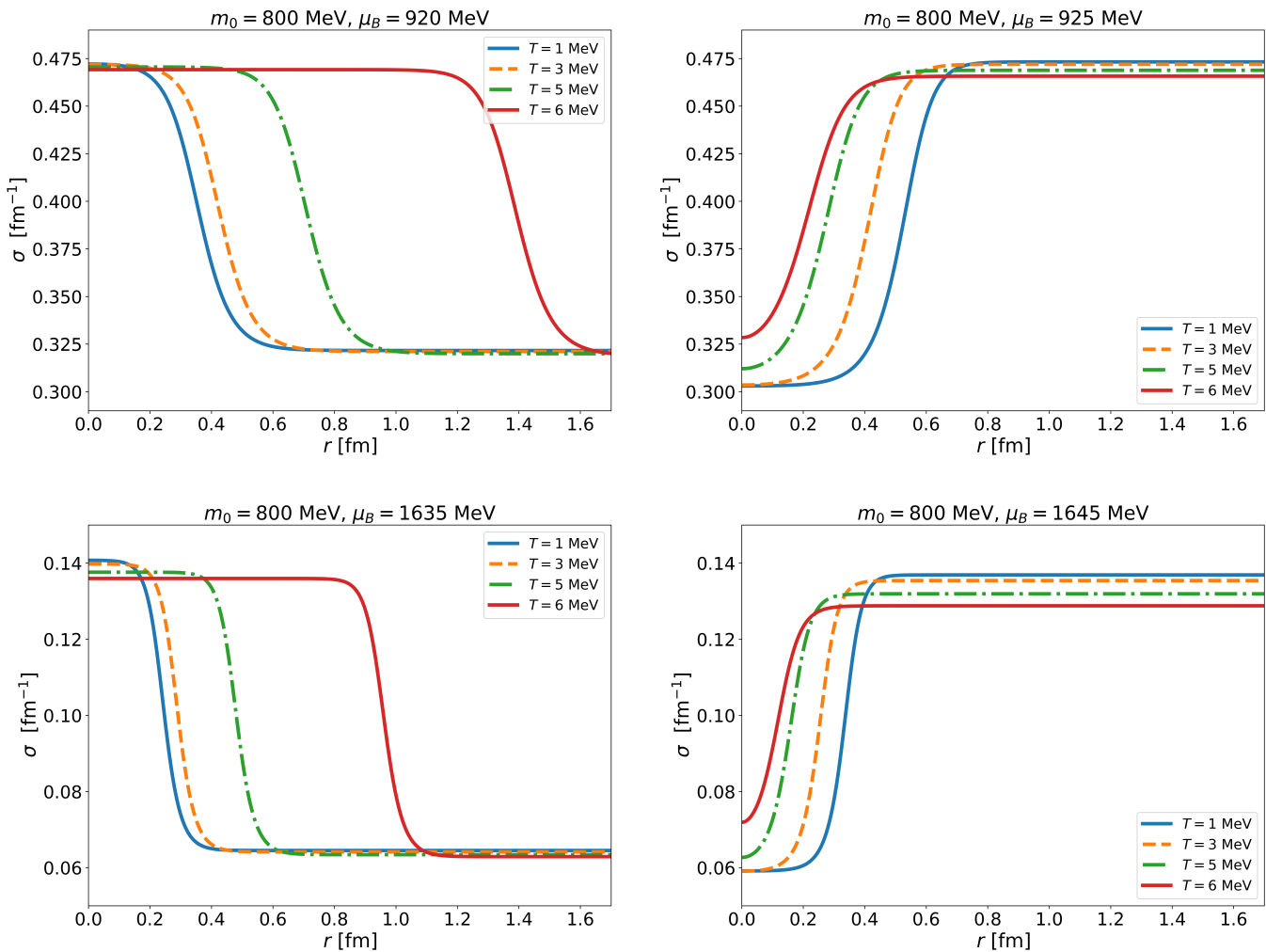}
\caption{Bounce solution $\sigma(r)$ as a function of radial distance for several values of baryon chemical potential $\mu_B$ at various temperatures. The upper panels correspond to the region near the liquid--gas transition, while the lower panels correspond to the region near the chiral transition.}
\label{sigma_profile}
\end{figure*}

The three-dimensional action $S_3$ is obtained by evaluating the energy functional on the bounce configuration:
\begin{align}
    S_3 = \int d^3r \left[\frac{1}{2}(\nabla\sigma)^2 + \Delta\Omega(\sigma, T, \mu_B)\right],
    \label{eq:S3}
\end{align}
where $\Delta\Omega(\sigma, T, \mu_B) = \Omega(\sigma_{{\rm TV}}, T, \mu_B) - \Omega(\sigma_{\rm FV}, T, \mu_B)$ is the thermodynamic potential measured relative to the false vacuum. The bounce solution $\sigma(r)$ describes the spatial profile of the $\sigma$ field inside a critical bubble---one that is precisely balanced between expansion and collapse. Assuming spherical symmetry, which minimizes the action for a single scalar field, the bounce configuration satisfies the equation of motion
\begin{align}
    \frac{d^2\sigma(r)}{dr^2} + \frac{2}{r}\frac{d\sigma(r)}{dr} = \frac{\partial\Omega(\sigma, T, \mu_B)}{\partial\sigma},
    \label{eq:bounce}
\end{align}
subject to the boundary conditions
\begin{align}
    \lim_{r \to \infty} \sigma(r) = \sigma_{\rm FV}, \qquad \left.\frac{d\sigma(r)}{dr}\right|_{r=0} = 0.
    \label{eq:BC}
\end{align}
The first condition ensures that the field approaches the false vacuum value far from the bubble center, reflecting the fact that the bubble is embedded in the metastable phase. The second condition enforces regularity at the origin, which is required for the solution to have finite energy. Physically, the bounce solution interpolates smoothly from a value near the true vacuum at the bubble center to the false vacuum at spatial infinity, with the bubble wall corresponding to the region of rapid field variation.

The bounce solutions $\sigma(r)$ obtained by solving Eq.~\eqref{eq:bounce} are shown in Fig.~\ref{sigma_profile} for several representative values of $\mu_B$ at various temperatures, with $m_0 = 800$ MeV. The structure of the bounce profile depends crucially on which side of the first-order phase transition line the system is located in the phase diagram (see Fig.~\ref{phase_dia}).
We first consider the cases with $\mu_B = 920$ MeV (near the liquid--gas transition) and $\mu_B = 1635$ MeV (near the chiral transition), which lie to the left of their respective first-order transition curves in Fig.~\ref{phase_dia}. In this region, the true vacuum corresponds to a larger value of $\sigma$, while the false vacuum has a smaller value. Taking $\mu_B = 920$ MeV as an example, the $\sigma$ field at the bubble center takes the true vacuum value of approximately $\sigma \approx 0.466~\mathrm{fm}^{-1}$ ($\approx 92$ MeV). As the $r$ increases, the field asymptotically approaches the false vacuum value $\sigma_{\rm FV} \approx 0.325~\mathrm{fm}^{-1}$ ($\approx 64$ MeV). A similar profile structure is observed for $\mu_B = 1635$ MeV, albeit with different numerical values for the true and false vacua.

\begin{figure}[htbp]
\centering
\includegraphics[width=1\hsize]{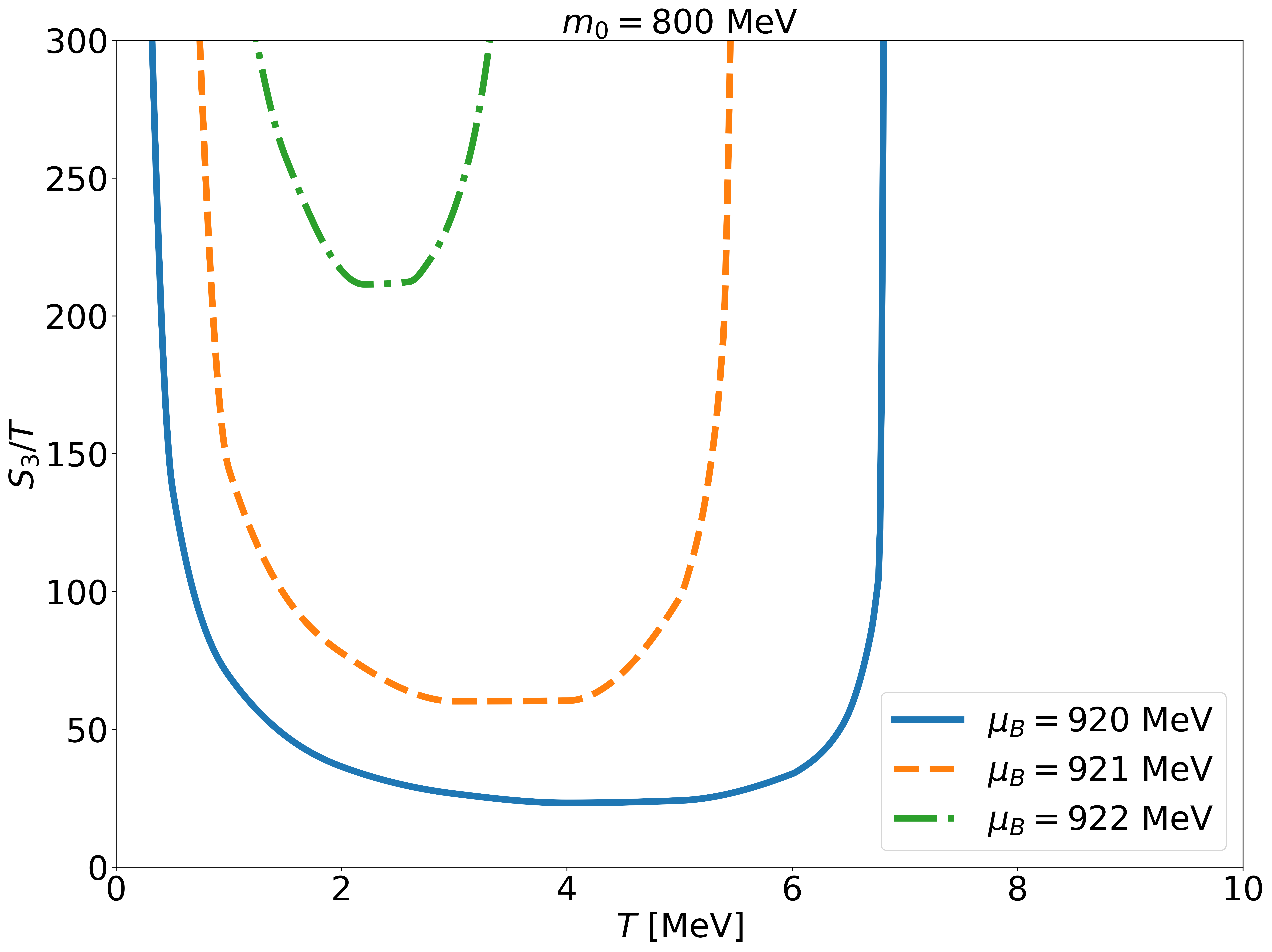}
\includegraphics[width=1\hsize]{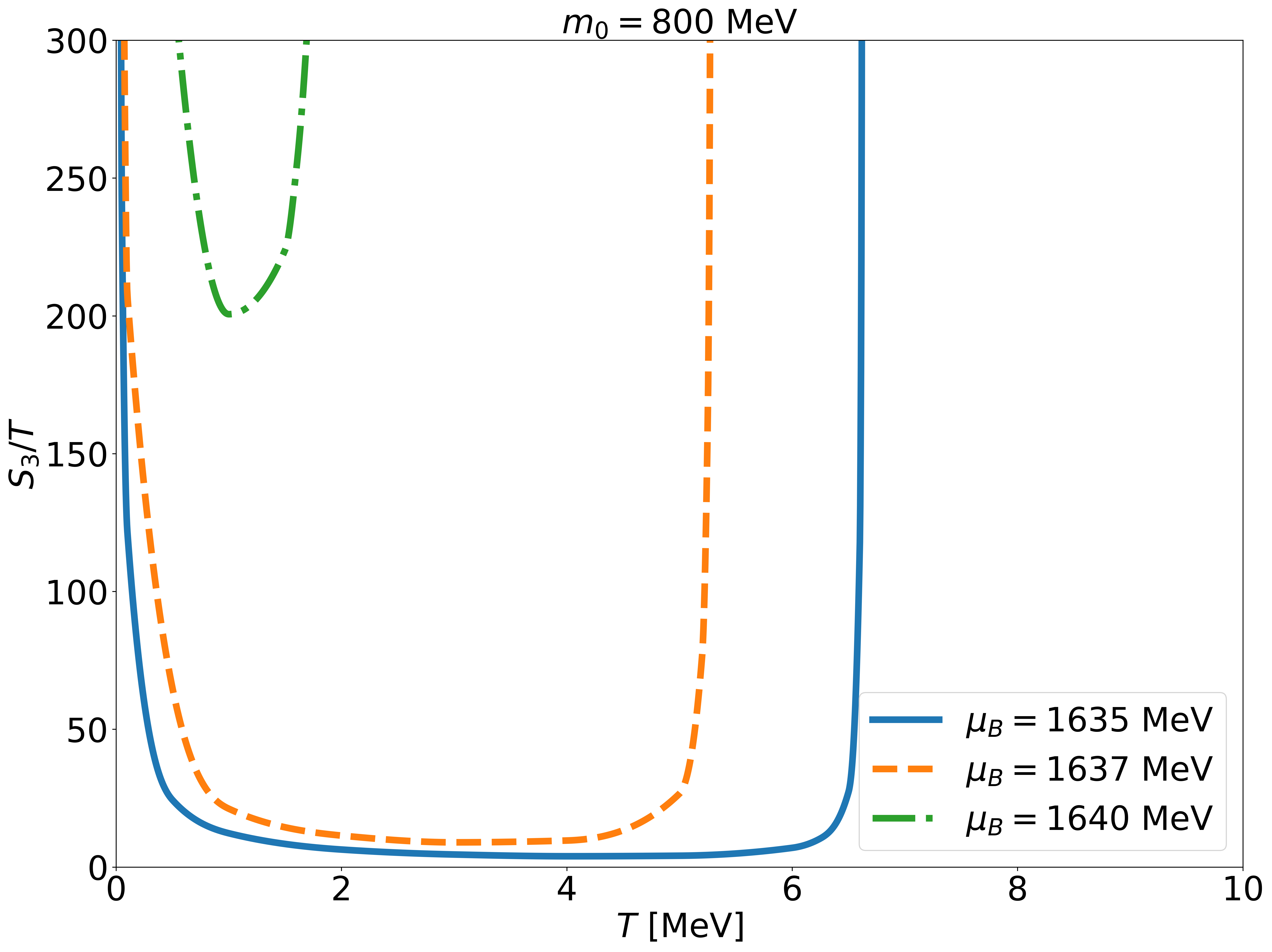}
\caption{$S_3/T$ as a function of temperature $T$ for several values of $\mu_B$ in the vicinity of the liquid–gas phase transition (upper panel) and the chiral phase transition (lower panel). }
\label{s3_lq_chi}
\end{figure}

In contrast, for $\mu_B = 925$ MeV and $\mu_B = 1645$ MeV, the system lies to the right of the first-order transition curves in Fig.~\ref{phase_dia}. In this case, the roles of the vacua are reversed: the true vacuum inside the bubble now corresponds to a smaller value of $\sigma$, while the false vacuum outside has a larger value. Consequently, the bounce profile exhibits an inverted structure compared to the previous cases, with $\sigma(r)$ increasing from the bubble center toward spatial infinity.
This reversal of the vacuum structure can be understood from the behavior of the thermodynamic potential across the phase transition. At the critical temperature $T_c$, the two local minima of the thermodynamic potential become degenerate with distinct $\sigma$ values. For temperatures below $T_c$ on the left side of the transition line, the minimum with larger $\sigma$ becomes the global minimum (true vacuum), while the minimum with smaller $\sigma$ becomes metastable (false vacuum). Crossing to the right side of the transition line reverses this hierarchy, making the smaller-$\sigma$ minimum the true vacuum. This exchange of stability between the two minima is reflected in the qualitatively different bounce profiles observed on either side of the first-order transition curve.

After the $\sigma$ profile is determined, from Eq.~(\ref{eq:S3}), we can obtain the $S_3 / T$ value for a determined $\mu_B$ and $T$. In the cosmological context, the early Universe cools down from high temperatures, and first-order phase transitions proceed through the nucleation of true vacuum bubbles within the metastable false vacuum. For a given temperature $T$ below the critical temperature $T_c$, the relevant configuration for bubble nucleation corresponds to the transition from the metastable high-temperature phase (false vacuum) to the stable low-temperature phase (true vacuum). As illustrated in Fig.~\ref{sigma_profile}, for $\mu_B$ values to the left of the first-order transition curve---such as $\mu_B = 920$ MeV for the liquid--gas transition and $\mu_B = 1635$ MeV for the chiral transition---the false vacuum corresponds to the lower-$\sigma$ phase characteristic of higher temperatures, while the true vacuum corresponds to the higher-$\sigma$ phase favored at lower temperatures. This is precisely the physical situation relevant for cosmological phase transitions: as the Universe cools below $T_c$, the system becomes trapped in the metastable false vacuum (smaller $\sigma$), and thermal fluctuations drive the nucleation of bubbles containing the true vacuum (larger $\sigma$). Consequently, we focus our calculation of the bounce action $S_3/T$ on this region, where the nucleation dynamics directly determines the phase transition rate $\beta/H$ and the resulting GW spectrum. The opposite case---$\mu_B$ values to the right of the transition curve---corresponds to a heating scenario rather than the cooling history of the early Universe, and is therefore not relevant for our purposes.

Figure~\ref{s3_lq_chi} presents the temperature dependence of the bounce action $S_3/T$ for several representative values of $\mu_B$. The upper panel displays results near the liquid--gas phase transition, while the lower panel shows the corresponding behavior near the chiral phase transition.
A characteristic ``U-shaped'' profile emerges in both cases: starting from low temperatures, $S_3/T$ initially decreases with increasing temperature, reaches a minimum at an intermediate temperature, and subsequently increases as the temperature approaches the critical temperature $T_c$. This non-monotonic behavior can be understood from the interplay between two competing effects. At temperatures well below $T_c$, the potential barrier between the false and true vacua is relatively high, resulting in a large bounce action. As the temperature increases toward $T_*$, thermal fluctuations become more effective at overcoming the barrier, and the potential difference between the two vacua decreases, leading to a reduction in $S_3/T$. However, as $T$ approaches $T_c$ from below, the two minima of the thermodynamic potential become increasingly degenerate. In this limit, the critical bubble radius $R_c$ diverges as the driving force for the phase transition vanishes, causing $S_3/T$ to diverge as well.
The behavior of $S_3/T$ has direct implications for the nucleation dynamics. Since the nucleation rate scales as $\Gamma \propto \exp(-S_3/T)$, a large value of $S_3/T$ exponentially suppresses bubble formation, leaving the system trapped in the metastable false vacuum. 

\section{GW spectra from first-order phase transition}\label{sec_4}

The GW spectrum from a first-order phase transition is governed mainly by two key parameters evaluated at the nucleation temperature $T_n$: the inverse duration of the transition $\beta/H$, which controls the peak frequency, and the transition strength $\alpha$, which determines the amplitude. In this section, we define these quantities within the framework of the parity doublet model at finite baryon chemical potential and then obtain the corresponding GWs.


The nucleation temperature $T_n$ is defined as the temperature at which the nucleation rate becomes comparable to the expansion rate of the Universe. As the Universe cools below the critical temperature $T_c$, the bounce action $S_3/T$ decreases from its divergent value at $T_c$, and bubble nucleation becomes increasingly probable. The nucleation condition is conventionally expressed as the requirement that approximately one bubble nucleates per Hubble volume per Hubble time~\cite{ellis2020gravitational,RN20},
\begin{equation}
\Gamma(t)/H^4 \sim 1,
\end{equation}
where $H$ is the Hubble parameter determined by the Friedmann equation
\begin{equation}
H = \sqrt{\frac{\varepsilon_T}{3m_p^2}},
\end{equation}
with $m_p = 2.435 \times 10^{18}$~GeV the reduced Planck mass and $\varepsilon_T$ the total energy density of the Universe at temperature $T$. Substituting the nucleation rate $\Gamma = A(T) e^{-S_3/T}$ into the above condition, the nucleation temperature $T_n$ can be determined numerically.

The first key parameter is the inverse duration of the phase transition, parametrized by the ratio $\beta/H$. This quantity measures how rapidly the transition completes relative to the Hubble expansion and is defined as
\begin{equation}
\frac{\beta}{H} = T_n \left. \frac{d(S_3/T)}{dT} \right|_{T_n}.
\end{equation}
A large $\beta/H$ corresponds to a fast transition producing many small bubbles, yielding a higher peak frequency, while a small $\beta/H$ indicates a slow transition with fewer, larger bubbles that generate a stronger GW signal at lower frequencies.

The second key parameter is the transition strength $\alpha$, which quantifies the energy available to source GWs. At finite baryon chemical potential, its definition requires careful treatment of the baryon number contribution. The latent heat released during the transition is given by the discontinuity in the trace of the energy-momentum tensor, $\Delta(\varepsilon - 3p)$, across the phase boundary. However, at finite $\mu_B$, the change in baryon number density between the two phases also contributes to the energy density difference. Since this contribution arises from the variation in rest mass energy and cannot be converted into kinetic energy of the plasma to drive GWs, it must be subtracted. We therefore define
\begin{equation}
\alpha = \frac{-\Delta(\varepsilon - \mu_B n_B) + 3\Delta p}{4\varepsilon_r},
\end{equation}
where $\Delta$ denotes the difference between the true and false vacuum values, and $\varepsilon_r$ is the radiation energy density of the background plasma evaluated at $T_n$, which serves as the normalization scale.



To obtain the GW spectrum, we need to further determine the bubble wall velocity $v_w$ and the efficiency factors with which the released vacuum energy is converted into different forms of kinetic energy. The bubble wall velocity is in general model dependent and can be determined through various approaches, including the Boltzmann equation~\cite{Wang:2024wcs}, numerical simulations~\cite{kurki1996bubble}, and holographic methods~\cite{bigazzi2021bubble,Chen:2022cgj}. For strong first-order phase transitions, the bubble wall typically become relativistic shortly after nucleation. In this regime, a widely adopted approximation is the Jouguet detonation velocity~\cite{PhysRevD.25.2074,Lerambert-Potin:2021ohy,espinosa2010energy}, which depends on $\alpha$ as
\begin{equation}
\label{vw}
v_w = v_J = \frac{\sqrt{1/3} + \sqrt{\alpha^2 + 2\alpha/3}}{1 + \alpha}.
\end{equation}

The vacuum energy released during the phase transition is distributed among several channels: the kinetic energy of bulk fluid motion, MHD turbulence, and the gradient energy of the scalar field, characterized by the efficiency factors $\kappa_v$, $\kappa_{\mathrm{tb}}$, and $\kappa_\phi$, respectively. For relativistic bubble walls, the scalar field gradient contribution $\kappa_\phi$ is negligibly small~\cite{RN69}, and the GW production is dominated by the bulk fluid motion and turbulence.

The efficiency factor $\kappa_v$, representing the fraction of vacuum energy converted into bulk kinetic energy of the plasma, takes different functional forms depending on the hydrodynamic mode of bubble expansion~\cite{espinosa2010energy,RN69,kamionkowski1994gravitational}. Following the numerical fits of Ref.~\cite{espinosa2010energy}, we introduce the following functions
\begin{align}
\kappa_A &= v_w^{6/5} \frac{6.9\,\alpha}{1.36 - 0.037\sqrt{\alpha} + \alpha}, \\
\kappa_B &= \frac{\alpha^{2/5}}{0.017 + (0.997 + \alpha)^{2/5}}, \\
\kappa_C &= \frac{\sqrt{\alpha}}{0.135 + \sqrt{0.98 + \alpha}}, \\
\kappa_D &= \frac{\alpha}{0.73 + 0.083\sqrt{\alpha} + \alpha}, \\
\delta\kappa &= -0.9\ln\!\left(\frac{\sqrt{\alpha}}{1 + \sqrt{\alpha}}\right).
\end{align}
The efficiency factor $\kappa_v$ is then given in different regions.
For subsonic deflagrations ($v_w < v_s$),
\begin{equation}
\kappa_v = \frac{v_s^{11/5}\,\kappa_A\,\kappa_B}{(v_s^{11/5} - v_w^{11/5})\,\kappa_B + v_w\,v_s^{6/5}\,\kappa_A}\,,
\end{equation}
for supersonic deflagrations ($v_s < v_w < v_J$),
\begin{equation}
\kappa_v = \kappa_B + (v_w - v_s)\,\delta\kappa + \frac{(v_w - v_s)^3}{(v_J - v_s)^3}\Big[\kappa_C - \kappa_B - (v_J - v_s)\,\delta\kappa\Big],
\end{equation}
and for detonations ($v_w \geq v_J$),
\begin{equation}
\kappa_v = \frac{(v_J - 1)^3\,(v_J/v_w)^{5/2}\,\kappa_C\,\kappa_D}{\left[(v_J - 1)^3 - (v_w - 1)^3\right]v_J^{5/2}\,\kappa_C + (v_w - 1)^3\,\kappa_D}\,,
\end{equation}
where $v_s = 1/\sqrt{3}$ is the speed of sound in the relativistic plasma. Since we adopt the Jouguet detonation velocity $v_w = v_J$ in Eq.~(\ref{vw}), the detonation branch applies throughout our analysis.
The efficiency factor for MHD turbulence is estimated from numerical simulations to be $\kappa_{\mathrm{tb}} = 0.05\,\kappa_v$, based on findings that the turbulent contribution amounts to roughly $5$--$10\%$ of the bulk kinetic energy~\cite{RN69,RN68}.

With all the phase transition parameters determined, we can now compute the GW spectrum. The total GW energy density spectrum receives contributions from three sources: sound waves in the plasma, MHD turbulence, and bubble wall collisions,
\begin{equation}
h^2\Omega_{\mathrm{GW}}(f) = h^2\Omega_{\mathrm{sw}}(f) + h^2\Omega_{\mathrm{tb}}(f) + h^2\Omega_{\mathrm{env}}(f).
\end{equation}

The dominant contribution arises from sound waves generated by the bulk fluid motion after bubble collisions. The corresponding spectrum, obtained from numerical simulations~\cite{RN15,RN69}, is given by
\begin{equation}\label{11}
h^2\Omega_{\mathrm{sw}} = 2.65\times10^{-6} \left(\frac{H}{\beta}\right) \left(\frac{\kappa_v\alpha}{1+\alpha}\right)^2 \left(\frac{100}{g}\right)^{\frac{1}{3}} v_w\,S_{\mathrm{sw}}(f),
\end{equation}
where $g$ denotes the number of relativistic degrees of freedom at $T_n$, and the spectral shape function is
\begin{equation}\label{13}
S_{\mathrm{sw}}(f) = \left(\frac{f}{f_{\mathrm{sw}}}\right)^3 \left(\frac{7}{4+3\left(\frac{f}{f_{\mathrm{sw}}}\right)^2}\right)^{\frac{7}{2}},
\end{equation}
with the peak frequency
\begin{equation}\label{15}
f_{\mathrm{sw}} = 1.9\times10^{-5}\,\frac{1}{v_w}\frac{\beta}{H}\frac{T_n}{100\,\mathrm{GeV}} \left(\frac{g}{100}\right)^{\frac{1}{6}}\,\mathrm{Hz}.
\end{equation}

The contribution from MHD turbulence takes a similar form~\cite{RN15,RN69},
\begin{equation}\label{12}
h^2\Omega_{\mathrm{tb}} = 3.35\times10^{-4} \left(\frac{H}{\beta}\right) \left(\frac{\kappa_{\mathrm{tb}}\alpha}{1+\alpha}\right)^2 \left(\frac{100}{g}\right)^{\frac{1}{3}} v_w\,S_{\mathrm{tb}}(f),
\end{equation}
with the spectral shape function
\begin{equation}\label{14}
S_{\mathrm{tb}}(f) = \left(\frac{f}{f_{\mathrm{tb}}}\right)^3 \left(1+\frac{f}{f_{\mathrm{tb}}}\right)^{-\frac{11}{3}} \left(1+\frac{8\pi f}{h_*}\right)^{-1},
\end{equation}
where
\begin{equation}
h_* = 1.65\times10^{-6}\,\frac{T_n}{100\,\mathrm{GeV}} \left(\frac{g}{100}\right)^{\frac{1}{6}}\,\mathrm{Hz}
\end{equation}
is the red-shifted Hubble rate at the transition epoch, and the peak frequency is
\begin{equation}\label{16}
f_{\mathrm{tb}} = 2.7\times10^{-5}\,\frac{1}{v_w}\frac{\beta}{H}\frac{T_n}{100\,\mathrm{GeV}} \left(\frac{g}{100}\right)^{\frac{1}{6}}\,\mathrm{Hz}.
\end{equation}

Finally, the contribution from direct bubble wall collisions is~\cite{RN69}
\begin{equation}
\begin{split}
h^2\Omega_{\mathrm{env}} = 1.67\times10^{-5} \left(\frac{H}{\beta}\right)^2 \left(\frac{\kappa_\phi\alpha}{1+\alpha}\right)^2 \\
\times \left(\frac{100}{g}\right)^{\frac{1}{3}} \left(\frac{0.11\,v_w^3}{0.42+v_w^2}\right) S_{\mathrm{env}}(f),
\end{split}
\end{equation}
with
\begin{equation}
S_{\mathrm{env}}(f) = \frac{3.8\,(f/f_{\mathrm{env}})^{2.8}}{1+2.8\,(f/f_{\mathrm{env}})^{3.8}}
\end{equation}
and
\begin{equation}
f_{\mathrm{env}} = 1.65\times10^{-5}\,\frac{0.62}{1.8-0.1\,v_w+v_w^2}\frac{T_n}{100\,\mathrm{GeV}} \left(\frac{g}{100}\right)^{\frac{1}{6}}\,\mathrm{Hz}.
\end{equation}
Here $\kappa_\phi$ denotes the fraction of vacuum energy converted into the gradient energy of the scalar field. For relativistic bubble walls, $\kappa_\phi$ is negligibly small compared to $\kappa_v$ and $\kappa_{\mathrm{tb}}$~\cite{RN69}, so that the GW spectrum is dominated by the sound wave and turbulence contributions.

\begin{table}[htbp]
    \centering
    \label{gwpa}
\begin{subtable}{\linewidth} 
        \centering
        \begin{tabular}{p{1.3cm}|p{1.3cm} p{1.3cm} p{1.3cm} p{1.3cm}}
\hline
\hline
$\mu$ [MeV] & 920 & 921 & 921.7 & 921.8  \\ 
\hline
$\alpha$ &  0.199 & 0.409 & 2.75 & 7.57 \\
$\beta/H$ &  33225 & 2428 &  353 & 46 \\
\hline
\hline
\end{tabular}
        \caption{Liquid-gas phase transition.}
        \label{lg}
    \end{subtable}
\vspace{2em}
 \begin{subtable}{\linewidth}
        \centering
        \begin{tabular}{p{1.3cm}|p{1.3cm} p{1.3cm} p{1.3cm} p{1.3cm}}
\hline
\hline
$\mu$ [MeV] & 1635& 1637& 1639.8& 1639.9\\ 
\hline
$\alpha\times10^{5}$&  6.37& 4.43& 1.45& 1.39\\
 $\beta/H$    &  53036& 17805&  227& 129\\
 \hline
  \hline
\end{tabular}
        \caption{Chiral phase transition.}
        \label{ch}
    \end{subtable}
    \caption{Parameters $\alpha$ and $\beta$ for several choices of chemical potential near two phase transition regions.}
\end{table}

\begin{figure*}
        \centering
    \includegraphics[width=1\linewidth]{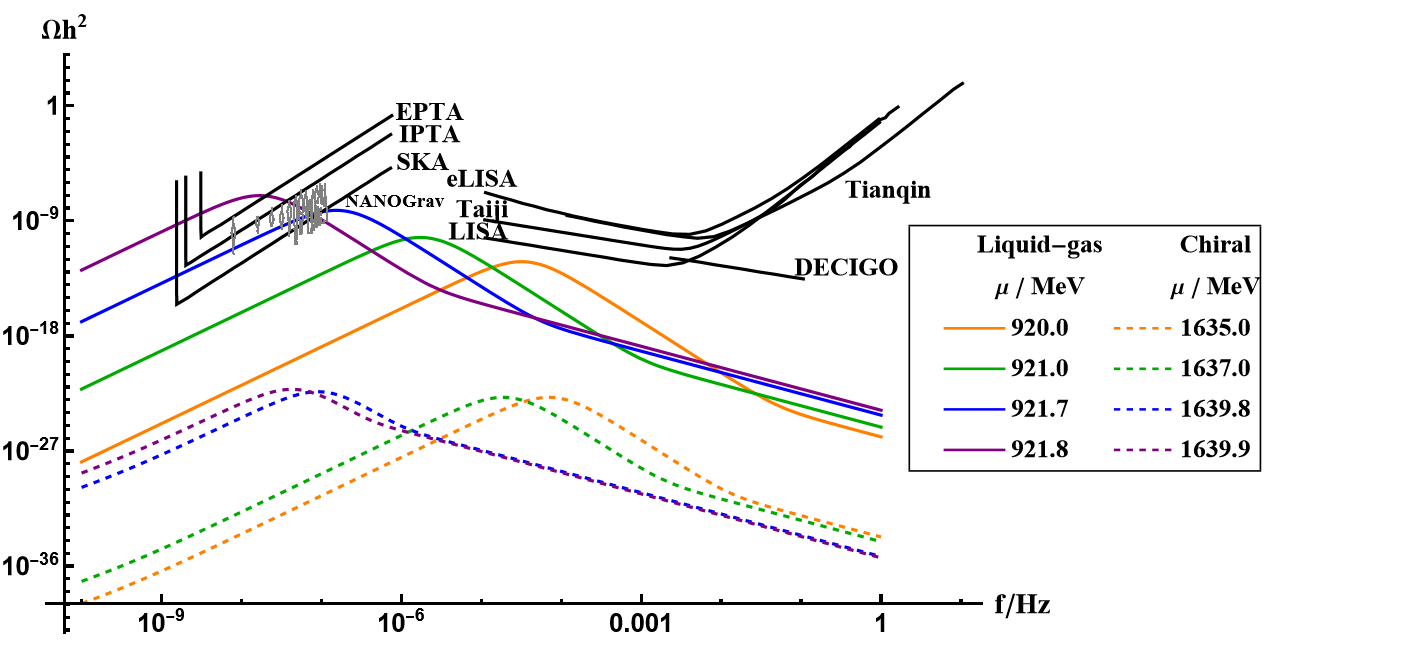}
    \caption{GW spectra with different chemical potential from the liquid-gas transition ( Solid curves) and the chiral transition (dashed curves).}
    \label{dpgw}  
\end{figure*}

The phase transition parameters $\alpha$ and $\beta/H$ for selected values of chemical potential are summarized in Tab.~\ref{lg} for the liquid--gas transition and Tab.~\ref{ch} for the chiral transition. In both cases, the inverse duration $\beta/H$ exhibits qualitatively similar behavior — it takes comparable values and decreases monotonically as the chemical potential approaches the endpoint of the first-order transition line. This decrease reflects the fact that, closer to the endpoint, the potential barrier between the two vacua becomes steeper and the transition dynamics slows down, resulting in fewer but larger bubbles.

The transition strength $\alpha$, however, exhibits different behavior in the two transitions. For the liquid--gas transition, $\alpha$ increases with chemical potential and reaches $O(1)$ values, signaling a strong first-order phase transition capable of producing significant GW signals. In contrast, for the chiral transition occurring at much higher chemical potential ($\mu_B \sim 1635$--$1640$ MeV), $\alpha$ is suppressed by approximately five orders of magnitude relative to the liquid--gas case. This dramatic suppression arises because the transition strength is normalized by the radiation energy density $\varepsilon_r$ and receives much larger contributions at high $\mu_B$, thereby diluting the relative importance of the latent heat released during the transition. Moreover, $\alpha$ decreases with increasing chemical potential in the chiral transition, opposite to the trend observed in the liquid--gas case. As a result, the two types of phase transitions produce GW signals of vastly different amplitudes, providing a clear observational signature to distinguish between two type of phase transitions.


In Fig.~\ref{dpgw}, we show the GW spectra for several  values of chemical potential, comparing the contributions from the liquid--gas transition (solid curves) and the chiral transition (dashed curves). For both transitions, the peak frequency shifts toward lower values as the chemical potential increases, which is a direct consequence of the decreasing $\beta/H$.
For the liquid--gas transition, the peak frequencies span a broad range down to the nanohertz band, depending on the choice of chemical potential.  Notably, in a narrow window near the endpoint of the first-order transition line (e.g., $\mu_B = 921.8$~MeV), where $\beta/H$ becomes sufficiently small and $\alpha$ is of order unity, the peak frequency enters the nanohertz regime accessible to PTAs and provides a good fit to the NANOGrav 15-year data, suggesting that a first-order nuclear liquid--gas transition in the early Universe could serve as a viable source for the observed stochastic GW background.
In contrast, the GW spectra from the chiral phase transition are much weaker, reflecting the strongly suppressed values of $\alpha$ discussed above. Even in the most favorable case near the critical point of the chiral transition, the signal amplitude remains well below the projected sensitivity curves of all current and planned detectors.

This  separation in signal strength indicate that the current and planned GW experiments is sufficient to detect signals only from the liquid--gas phase transition occurring at relatively low baryon densities. For phase transitions taking place at high baryon densities, such as the first-order chiral phase transition or a potential first-order deconfinement transition, the background energy density of the Universe at the corresponding epoch is substantially larger, which greatly suppresses the transition strength $\alpha$ and consequently the amplitude of the resulting GW spectrum. This suppression renders the GW signals from high-density QCD phase transitions far too weak to be observed with foreseeable detector sensitivities, posing a fundamental challenge for probing the high-density region of the QCD phase diagram through GW astronomy.

Finally, we note that in our approach, we have identified two distinct types of first-order phase transitions---the liquid-gas and the chiral phase transition---and computed their corresponding GW spectra separately. However, from the observational perspective, the inverse problem is considerably more challenging: a detected GW spectrum alone cannot unambiguously determine the microscopic nature of the underlying first-order phase transition, whether it originates from a liquid-gas transition, a chiral transition, or a potential color deconfinement transition. This degeneracy arises because the GW spectrum is primarily characterized by two macroscopic parameters, $\alpha$ and $\beta/H$, which encode the transition dynamics but do not directly reveal the specific order parameter driving the transition. Nevertheless, our results suggest that the amplitude of the GW signal provides a useful diagnostic: phase transitions occurring at high baryon densities, where the background energy density of the Universe is substantially larger, produce GW signals with strongly suppressed amplitudes due to the dilution of the transition strength $\alpha$. Therefore, a strong GW signal in the nanohertz band would favor a phase transition at relatively low baryon density, such as the liquid--gas transition, while the absence of detectable signals cannot exclude the occurrence of phase transitions at higher densities. Combining GW observations with complementary probes---such as neutron star observations, heavy-ion collision experiments, and future measurements of the QCD equation of state---will be essential for breaking this degeneracy and identifying the nature of the underlying phase transition.

\section{Summary and Discussion}\label{sec-summary}

In this work, we have investigated the GW spectrum generated by first-order QCD phase transitions within the framework of the parity doublet model. The parity doublet model provides a unique description of baryonic matter by incorporating the chiral invariant mass $m_0$, which represents the portion of nucleon mass that persists even when chiral symmetry is restored. By fixing $m_0 = 800$~MeV and fitting the remaining model parameters to nuclear saturation properties, we have systematically studied the bubble nucleation dynamics and the resulting GW spectra from both the nuclear liquid--gas phase transition and the chiral phase transition at finite baryon chemical potential.

The low-temperature phase diagram of the parity doublet model exhibits two distinct first-order phase transition regions: the liquid--gas transition at relatively low baryon chemical potential ($\mu_B \sim 920$~MeV) and the chiral phase transition at higher chemical potential ($\mu_B \sim 1635$~MeV). For each transition, we have solved the bounce equation to obtain the critical bubble profiles and computed the three-dimensional Euclidean action $S_3/T$ as a function of temperature. The bounce action displays a characteristic U-shaped temperature dependence, reflecting the competition between the potential barrier height and the degeneracy of the two vacua near the critical temperature.

The GW spectra from these two types of phase transitions exhibit  different characteristics. For the liquid--gas phase transition, the transition strength $\alpha$ can reach $\mathcal{O}(1)$ values near the endpoint of the first-order transition line, and the inverse duration $\beta/H$ can be reduced to $\mathcal{O}(10)$--$\mathcal{O}(100)$. This combination yields GW signals with peak frequencies spanning from the millihertz band down to the nanohertz regime. In particular, for $\mu_B$ around 921.8 $\,$ MeV, the predicted spectrum provides a good fit to the NANOGrav 15-year data, suggesting that a first-order nuclear liquid--gas transition in the early Universe, facilitated by a large initial baryon asymmetry and subsequent dilution through little inflation, could serve as a viable source for the observed stochastic GW background. In contrast, the chiral phase transition occurring at much higher chemical potential produces GW signals that are suppressed by approximately five orders of magnitude in $\alpha$ relative to the liquid--gas case, rendering them well below the sensitivity of all current and planned detectors.

Several aspects of this work merit further discussion and point to promising directions for future investigation.
First, the chiral invariant mass $m_0$ plays a central role in determining the phase structure and, consequently, the GW spectrum. In this study, we have fixed $m_0 = 800$~MeV as a representative value consistent with neutron star observations. However, varying $m_0$ modifies the location and strength of the first-order phase transitions in the $T$--$\mu_B$ plane~\cite{Motohiro:2015taa,Marczenko:2025kpv}, which in turn alters the nucleation dynamics, the transition parameters $\alpha$ and $\beta/H$, and ultimately the peak frequency and amplitude of the GW spectrum. Since the chiral invariant mass directly quantifies the contribution to nucleon mass that does not originate from spontaneous chiral symmetry breaking, the GW signal carries an imprint of this fundamental quantity. Therefore, future GW observations by pulsar timing arrays and space-based interferometers could, in principle, provide complementary constraints on $m_0$ and shed light on the origin of nucleon mass from a cosmological perspective. A systematic study of the $m_0$ dependence of the GW spectrum, combined with constraints from neutron star observations from NICER and GW detections from LIGO/Virgo/KAGRA, would establish a novel multi-messenger approach to understanding the origin of hadron mass.

Second, within the present framework based on the parity doublet model, 
the inverse duration parameter of the phase transition, $\beta/H$, 
is typically large, $\mathcal{O}(10^{2})$--$\mathcal{O}(10^{4})$, 
even in the most favorable region near the endpoint of the first-order 
transition curve. Such values are compatible with the production 
of stochastic GW backgrounds potentially detectable by 
pulsar timing arrays and future space-based interferometers in the 
liquid--gas transition scenario. However, they remain parametrically 
far from the slow-transition regime, $\beta/H \sim \mathcal{O}(1)$--$\mathcal{O}(10)$, 
which is generally required for efficient primordial black hole formation 
\cite{Hashino:2021qoq,Lewicki:2023ioy,Gouttenoire:2023naa}. 
Therefore, while GW signals may be observable, 
primordial black hole production is not expected within the present setup. It would be interesting to explore extensions of the parity doublet model that could naturally realize the small value of $\beta / H$. One promising direction is the introduction of a dilaton field~\cite{Zhang:2025kbu,Jiang:2025ofd} that couples to the chiral condensate, which could provide an additional scalar degree of freedom whose dynamics governs the phase transition. In such an extended framework, the interplay between the dilaton potential and the chiral mean field could potentially realizing the ultra-supercooling regime with sufficiently small $\beta/H$ and open the possibility of connecting the QCD phase transition in the parity doublet model to primordial black hole formation and dark matter production, establishing a link between the microscopic physics of hadron mass generation and the macroscopic content of the Universe.

Third, we note that our analysis has been performed within the mean-field approximation, and several refinements could be incorporated in future work. These include the effects of quantum and thermal fluctuations beyond mean field, the inclusion of strange quarks and hyperons within the $SU(3)$  parity doublet model, and a more detailed treatment of the bubble wall velocity beyond the Jouguet detonation approximation. The extension to the $SU(3)$ sector is particularly relevant, as the appearance of strangeness-carrying baryons could modify the phase structure at high density and potentially introduce additional first-order phase transitions that contribute to the GW spectrum.

In conclusion, the parity doublet model provides a physically motivated and quantitatively tractable framework for studying GW production from first-order QCD phase transitions. The distinct signatures from the liquid--gas and chiral transitions offer observational handles to probe different density regimes of the QCD phase diagram. The connection between the chiral invariant mass $m_0$ and the GW spectrum establishes a novel pathway to investigate the origin of nucleon mass through GW astronomy, complementing ongoing efforts from neutron star observations and heavy-ion collision experiments.


\bibliography{ref_PDM_2026.bib}

\begin{thebibliography}{72}%
\makeatletter
\providecommand \@ifxundefined [1]{%
 \@ifx{#1\undefined}
}%
\providecommand \@ifnum [1]{%
 \ifnum #1\expandafter \@firstoftwo
 \else \expandafter \@secondoftwo
 \fi
}%
\providecommand \@ifx [1]{%
 \ifx #1\expandafter \@firstoftwo
 \else \expandafter \@secondoftwo
 \fi
}%
\providecommand \natexlab [1]{#1}%
\providecommand \enquote  [1]{``#1''}%
\providecommand \bibnamefont  [1]{#1}%
\providecommand \bibfnamefont [1]{#1}%
\providecommand \citenamefont [1]{#1}%
\providecommand \href@noop [0]{\@secondoftwo}%
\providecommand \href [0]{\begingroup \@sanitize@url \@href}%
\providecommand \@href[1]{\@@startlink{#1}\@@href}%
\providecommand \@@href[1]{\endgroup#1\@@endlink}%
\providecommand \@sanitize@url [0]{\catcode `\\12\catcode `\$12\catcode
  `\&12\catcode `\#12\catcode `\^12\catcode `\_12\catcode `\%12\relax}%
\providecommand \@@startlink[1]{}%
\providecommand \@@endlink[0]{}%
\providecommand \url  [0]{\begingroup\@sanitize@url \@url }%
\providecommand \@url [1]{\endgroup\@href {#1}{\urlprefix }}%
\providecommand \urlprefix  [0]{URL }%
\providecommand \Eprint [0]{\href }%
\providecommand \doibase [0]{https://doi.org/}%
\providecommand \selectlanguage [0]{\@gobble}%
\providecommand \bibinfo  [0]{\@secondoftwo}%
\providecommand \bibfield  [0]{\@secondoftwo}%
\providecommand \translation [1]{[#1]}%
\providecommand \BibitemOpen [0]{}%
\providecommand \bibitemStop [0]{}%
\providecommand \bibitemNoStop [0]{.\EOS\space}%
\providecommand \EOS [0]{\spacefactor3000\relax}%
\providecommand \BibitemShut  [1]{\csname bibitem#1\endcsname}%
\let\auto@bib@innerbib\@empty
\bibitem [{\citenamefont {Agazie}\ \emph {et~al.}(2023)\citenamefont {Agazie}
  \emph {et~al.}}]{NANOGrav:2023hde}%
  \BibitemOpen
  \bibfield  {author} {\bibinfo {author} {\bibfnamefont {G.}~\bibnamefont
  {Agazie}} \emph {et~al.} (\bibinfo {collaboration} {NANOGrav}),\ }\bibfield
  {title} {\bibinfo {title} {{The NANOGrav 15 yr Data Set: Observations and
  Timing of 68 Millisecond Pulsars}},\ }\href
  {https://doi.org/10.3847/2041-8213/acda9a} {\bibfield  {journal} {\bibinfo
  {journal} {Astrophys. J. Lett.}\ }\textbf {\bibinfo {volume} {951}},\
  \bibinfo {pages} {L9} (\bibinfo {year} {2023})},\ \Eprint
  {https://arxiv.org/abs/2306.16217} {arXiv:2306.16217 [astro-ph.HE]}
  \BibitemShut {NoStop}%
\bibitem [{\citenamefont {Lerambert-Potin}\ and\ \citenamefont
  {de~Freitas~Pacheco}(2021)}]{Lerambert-Potin:2021ohy}%
  \BibitemOpen
  \bibfield  {author} {\bibinfo {author} {\bibfnamefont {P.}~\bibnamefont
  {Lerambert-Potin}}\ and\ \bibinfo {author} {\bibfnamefont {J.~A.}\
  \bibnamefont {de~Freitas~Pacheco}},\ }\bibfield  {title} {\bibinfo {title}
  {{Gravitational Waves from the Cosmological Quark-Hadron Phase Transition
  Revisited}},\ }\href {https://doi.org/10.3390/universe7080304} {\bibfield
  {journal} {\bibinfo  {journal} {Universe}\ }\textbf {\bibinfo {volume} {7}},\
  \bibinfo {pages} {304} (\bibinfo {year} {2021})},\ \Eprint
  {https://arxiv.org/abs/2108.10727} {arXiv:2108.10727 [hep-ph]} \BibitemShut
  {NoStop}%
\bibitem [{\citenamefont {Zic}\ \emph {et~al.}(2023)\citenamefont {Zic} \emph
  {et~al.}}]{Zic:2023gta}%
  \BibitemOpen
  \bibfield  {author} {\bibinfo {author} {\bibfnamefont {A.}~\bibnamefont
  {Zic}} \emph {et~al.},\ }\bibfield  {title} {\bibinfo {title} {{The Parkes
  Pulsar Timing Array third data release}},\ }\href
  {https://doi.org/10.1017/pasa.2023.36} {\bibfield  {journal} {\bibinfo
  {journal} {Publ. Astron. Soc. Austral.}\ }\textbf {\bibinfo {volume} {40}},\
  \bibinfo {pages} {e049} (\bibinfo {year} {2023})},\ \Eprint
  {https://arxiv.org/abs/2306.16230} {arXiv:2306.16230 [astro-ph.HE]}
  \BibitemShut {NoStop}%
\bibitem [{\citenamefont {Reardon}\ \emph {et~al.}(2023)\citenamefont {Reardon}
  \emph {et~al.}}]{Reardon:2023gzh}%
  \BibitemOpen
  \bibfield  {author} {\bibinfo {author} {\bibfnamefont {D.~J.}\ \bibnamefont
  {Reardon}} \emph {et~al.},\ }\bibfield  {title} {\bibinfo {title} {{Search
  for an Isotropic Gravitational-wave Background with the Parkes Pulsar Timing
  Array}},\ }\href {https://doi.org/10.3847/2041-8213/acdd02} {\bibfield
  {journal} {\bibinfo  {journal} {Astrophys. J. Lett.}\ }\textbf {\bibinfo
  {volume} {951}},\ \bibinfo {pages} {L6} (\bibinfo {year} {2023})},\ \Eprint
  {https://arxiv.org/abs/2306.16215} {arXiv:2306.16215 [astro-ph.HE]}
  \BibitemShut {NoStop}%
\bibitem [{\citenamefont {Antoniadis}\ \emph
  {et~al.}(2023{\natexlab{a}})\citenamefont {Antoniadis} \emph
  {et~al.}}]{EPTA:2023sfo}%
  \BibitemOpen
  \bibfield  {author} {\bibinfo {author} {\bibfnamefont {J.}~\bibnamefont
  {Antoniadis}} \emph {et~al.} (\bibinfo {collaboration} {EPTA}),\ }\bibfield
  {title} {\bibinfo {title} {{The second data release from the European Pulsar
  Timing Array - I. The dataset and timing analysis}},\ }\href
  {https://doi.org/10.1051/0004-6361/202346841} {\bibfield  {journal} {\bibinfo
   {journal} {Astron. Astrophys.}\ }\textbf {\bibinfo {volume} {678}},\
  \bibinfo {pages} {A48} (\bibinfo {year} {2023}{\natexlab{a}})},\ \Eprint
  {https://arxiv.org/abs/2306.16224} {arXiv:2306.16224 [astro-ph.HE]}
  \BibitemShut {NoStop}%
\bibitem [{\citenamefont {Antoniadis}\ \emph
  {et~al.}(2023{\natexlab{b}})\citenamefont {Antoniadis} \emph
  {et~al.}}]{EPTA:2023fyk}%
  \BibitemOpen
  \bibfield  {author} {\bibinfo {author} {\bibfnamefont {J.}~\bibnamefont
  {Antoniadis}} \emph {et~al.} (\bibinfo {collaboration} {EPTA, InPTA:}),\
  }\bibfield  {title} {\bibinfo {title} {{The second data release from the
  European Pulsar Timing Array - III. Search for gravitational wave signals}},\
  }\href {https://doi.org/10.1051/0004-6361/202346844} {\bibfield  {journal}
  {\bibinfo  {journal} {Astron. Astrophys.}\ }\textbf {\bibinfo {volume}
  {678}},\ \bibinfo {pages} {A50} (\bibinfo {year} {2023}{\natexlab{b}})},\
  \Eprint {https://arxiv.org/abs/2306.16214} {arXiv:2306.16214 [astro-ph.HE]}
  \BibitemShut {NoStop}%
\bibitem [{\citenamefont {Xu}\ \emph {et~al.}(2023)\citenamefont {Xu} \emph
  {et~al.}}]{Xu:2023wog}%
  \BibitemOpen
  \bibfield  {author} {\bibinfo {author} {\bibfnamefont {H.}~\bibnamefont {Xu}}
  \emph {et~al.},\ }\bibfield  {title} {\bibinfo {title} {{Searching for the
  Nano-Hertz Stochastic Gravitational Wave Background with the Chinese Pulsar
  Timing Array Data Release I}},\ }\href
  {https://doi.org/10.1088/1674-4527/acdfa5} {\bibfield  {journal} {\bibinfo
  {journal} {Res. Astron. Astrophys.}\ }\textbf {\bibinfo {volume} {23}},\
  \bibinfo {pages} {075024} (\bibinfo {year} {2023})},\ \Eprint
  {https://arxiv.org/abs/2306.16216} {arXiv:2306.16216 [astro-ph.HE]}
  \BibitemShut {NoStop}%
\bibitem [{\citenamefont {Xue}\ \emph {et~al.}(2021)\citenamefont {Xue} \emph
  {et~al.}}]{Xue:2021gyq}%
  \BibitemOpen
  \bibfield  {author} {\bibinfo {author} {\bibfnamefont {X.}~\bibnamefont
  {Xue}} \emph {et~al.},\ }\bibfield  {title} {\bibinfo {title} {{Constraining
  Cosmological Phase Transitions with the Parkes Pulsar Timing Array}},\ }\href
  {https://doi.org/10.1103/PhysRevLett.127.251303} {\bibfield  {journal}
  {\bibinfo  {journal} {Phys. Rev. Lett.}\ }\textbf {\bibinfo {volume} {127}},\
  \bibinfo {pages} {251303} (\bibinfo {year} {2021})},\ \Eprint
  {https://arxiv.org/abs/2110.03096} {arXiv:2110.03096 [astro-ph.CO]}
  \BibitemShut {NoStop}%
\bibitem [{\citenamefont {Pisarski}\ and\ \citenamefont
  {Wilczek}(1984)}]{Pisarski:1983ms}%
  \BibitemOpen
  \bibfield  {author} {\bibinfo {author} {\bibfnamefont {R.~D.}\ \bibnamefont
  {Pisarski}}\ and\ \bibinfo {author} {\bibfnamefont {F.}~\bibnamefont
  {Wilczek}},\ }\bibfield  {title} {\bibinfo {title} {{Remarks on the Chiral
  Phase Transition in Chromodynamics}},\ }\href
  {https://doi.org/10.1103/PhysRevD.29.338} {\bibfield  {journal} {\bibinfo
  {journal} {Phys. Rev. D}\ }\textbf {\bibinfo {volume} {29}},\ \bibinfo
  {pages} {338} (\bibinfo {year} {1984})}\BibitemShut {NoStop}%
\bibitem [{\citenamefont {Svetitsky}\ and\ \citenamefont
  {Yaffe}(1982)}]{Svetitsky:1982gs}%
  \BibitemOpen
  \bibfield  {author} {\bibinfo {author} {\bibfnamefont {B.}~\bibnamefont
  {Svetitsky}}\ and\ \bibinfo {author} {\bibfnamefont {L.~G.}\ \bibnamefont
  {Yaffe}},\ }\bibfield  {title} {\bibinfo {title} {{Critical Behavior at
  Finite Temperature Confinement Transitions}},\ }\href
  {https://doi.org/10.1016/0550-3213(82)90172-9} {\bibfield  {journal}
  {\bibinfo  {journal} {Nucl. Phys. B}\ }\textbf {\bibinfo {volume} {210}},\
  \bibinfo {pages} {423} (\bibinfo {year} {1982})}\BibitemShut {NoStop}%
\bibitem [{\citenamefont {Yu}\ \emph {et~al.}(2014)\citenamefont {Yu},
  \citenamefont {Liu},\ and\ \citenamefont {Huang}}]{Yu:2014sla}%
  \BibitemOpen
  \bibfield  {author} {\bibinfo {author} {\bibfnamefont {L.}~\bibnamefont
  {Yu}}, \bibinfo {author} {\bibfnamefont {H.}~\bibnamefont {Liu}},\ and\
  \bibinfo {author} {\bibfnamefont {M.}~\bibnamefont {Huang}},\ }\bibfield
  {title} {\bibinfo {title} {{Spontaneous generation of local CP violation and
  inverse magnetic catalysis}},\ }\href
  {https://doi.org/10.1103/PhysRevD.90.074009} {\bibfield  {journal} {\bibinfo
  {journal} {Phys. Rev. D}\ }\textbf {\bibinfo {volume} {90}},\ \bibinfo
  {pages} {074009} (\bibinfo {year} {2014})},\ \Eprint
  {https://arxiv.org/abs/1404.6969} {arXiv:1404.6969 [hep-ph]} \BibitemShut
  {NoStop}%
\bibitem [{\citenamefont {Shao}(2023)}]{Shao:2023gho}%
  \BibitemOpen
  \bibfield  {author} {\bibinfo {author} {\bibfnamefont {S.-H.}\ \bibnamefont
  {Shao}},\ }\bibfield  {title} {\bibinfo {title} {{What's Done Cannot Be
  Undone: TASI Lectures on Non-Invertible Symmetries}},\ }in\ \href@noop {}
  {\emph {\bibinfo {booktitle} {{Theoretical Advanced Study Institute in
  Elementary Particle Physics 2023}: {Aspects of Symmetry}}}}\ (\bibinfo {year}
  {2023})\ \Eprint {https://arxiv.org/abs/2308.00747} {arXiv:2308.00747
  [hep-th]} \BibitemShut {NoStop}%
\bibitem [{\citenamefont {Fukushima}\ and\ \citenamefont
  {Hatsuda}(2011)}]{Fukushima:2010bq}%
  \BibitemOpen
  \bibfield  {author} {\bibinfo {author} {\bibfnamefont {K.}~\bibnamefont
  {Fukushima}}\ and\ \bibinfo {author} {\bibfnamefont {T.}~\bibnamefont
  {Hatsuda}},\ }\bibfield  {title} {\bibinfo {title} {{The phase diagram of
  dense QCD}},\ }\href {https://doi.org/10.1088/0034-4885/74/1/014001}
  {\bibfield  {journal} {\bibinfo  {journal} {Rept. Prog. Phys.}\ }\textbf
  {\bibinfo {volume} {74}},\ \bibinfo {pages} {014001} (\bibinfo {year}
  {2011})},\ \Eprint {https://arxiv.org/abs/1005.4814} {arXiv:1005.4814
  [hep-ph]} \BibitemShut {NoStop}%
\bibitem [{\citenamefont {Stephanov}(2004)}]{Stephanov:2004wx}%
  \BibitemOpen
  \bibfield  {author} {\bibinfo {author} {\bibfnamefont {M.~A.}\ \bibnamefont
  {Stephanov}},\ }\bibfield  {title} {\bibinfo {title} {{QCD Phase Diagram and
  the Critical Point}},\ }\href {https://doi.org/10.1143/PTPS.153.139}
  {\bibfield  {journal} {\bibinfo  {journal} {Prog. Theor. Phys. Suppl.}\
  }\textbf {\bibinfo {volume} {153}},\ \bibinfo {pages} {139} (\bibinfo {year}
  {2004})},\ \Eprint {https://arxiv.org/abs/hep-ph/0402115}
  {arXiv:hep-ph/0402115} \BibitemShut {NoStop}%
\bibitem [{\citenamefont {Affleck}\ and\ \citenamefont
  {Dine}(1985)}]{Affleck:1984fy}%
  \BibitemOpen
  \bibfield  {author} {\bibinfo {author} {\bibfnamefont {I.}~\bibnamefont
  {Affleck}}\ and\ \bibinfo {author} {\bibfnamefont {M.}~\bibnamefont {Dine}},\
  }\bibfield  {title} {\bibinfo {title} {{A New Mechanism for Baryogenesis}},\
  }\href {https://doi.org/10.1016/0550-3213(85)90021-5} {\bibfield  {journal}
  {\bibinfo  {journal} {Nucl. Phys. B}\ }\textbf {\bibinfo {volume} {249}},\
  \bibinfo {pages} {361} (\bibinfo {year} {1985})}\BibitemShut {NoStop}%
\bibitem [{\citenamefont {Dine}\ and\ \citenamefont
  {Kusenko}(2003)}]{Dine:2003ax}%
  \BibitemOpen
  \bibfield  {author} {\bibinfo {author} {\bibfnamefont {M.}~\bibnamefont
  {Dine}}\ and\ \bibinfo {author} {\bibfnamefont {A.}~\bibnamefont {Kusenko}},\
  }\bibfield  {title} {\bibinfo {title} {{The Origin of the matter - antimatter
  asymmetry}},\ }\href {https://doi.org/10.1103/RevModPhys.76.1} {\bibfield
  {journal} {\bibinfo  {journal} {Rev. Mod. Phys.}\ }\textbf {\bibinfo {volume}
  {76}},\ \bibinfo {pages} {1} (\bibinfo {year} {2003})},\ \Eprint
  {https://arxiv.org/abs/hep-ph/0303065} {arXiv:hep-ph/0303065} \BibitemShut
  {NoStop}%
\bibitem [{\citenamefont {Borghini}\ \emph {et~al.}(2000)\citenamefont
  {Borghini}, \citenamefont {Cottingham},\ and\ \citenamefont
  {Vinh~Mau}}]{Borghini:2000yp}%
  \BibitemOpen
  \bibfield  {author} {\bibinfo {author} {\bibfnamefont {N.}~\bibnamefont
  {Borghini}}, \bibinfo {author} {\bibfnamefont {W.~N.}\ \bibnamefont
  {Cottingham}},\ and\ \bibinfo {author} {\bibfnamefont {R.}~\bibnamefont
  {Vinh~Mau}},\ }\bibfield  {title} {\bibinfo {title} {{Possible cosmological
  implications of the quark hadron phase transition}},\ }\href
  {https://doi.org/10.1088/0954-3899/26/6/302} {\bibfield  {journal} {\bibinfo
  {journal} {J. Phys. G}\ }\textbf {\bibinfo {volume} {26}},\ \bibinfo {pages}
  {771} (\bibinfo {year} {2000})},\ \Eprint
  {https://arxiv.org/abs/hep-ph/0001284} {arXiv:hep-ph/0001284} \BibitemShut
  {NoStop}%
\bibitem [{\citenamefont {Boeckel}\ and\ \citenamefont
  {Schaffner-Bielich}(2010)}]{Boeckel:2009ej}%
  \BibitemOpen
  \bibfield  {author} {\bibinfo {author} {\bibfnamefont {T.}~\bibnamefont
  {Boeckel}}\ and\ \bibinfo {author} {\bibfnamefont {J.}~\bibnamefont
  {Schaffner-Bielich}},\ }\bibfield  {title} {\bibinfo {title} {{A little
  inflation in the early universe at the QCD phase transition}},\ }\href
  {https://doi.org/10.1103/PhysRevLett.105.041301} {\bibfield  {journal}
  {\bibinfo  {journal} {Phys. Rev. Lett.}\ }\textbf {\bibinfo {volume} {105}},\
  \bibinfo {pages} {041301} (\bibinfo {year} {2010})},\ \bibinfo {note}
  {[Erratum: Phys.Rev.Lett. 106, 069901 (2011)]},\ \Eprint
  {https://arxiv.org/abs/0906.4520} {arXiv:0906.4520 [astro-ph.CO]}
  \BibitemShut {NoStop}%
\bibitem [{\citenamefont {Boeckel}\ \emph {et~al.}(2011)\citenamefont
  {Boeckel}, \citenamefont {Schettler},\ and\ \citenamefont
  {Schaffner-Bielich}}]{Boeckel:2010bey}%
  \BibitemOpen
  \bibfield  {author} {\bibinfo {author} {\bibfnamefont {T.}~\bibnamefont
  {Boeckel}}, \bibinfo {author} {\bibfnamefont {S.}~\bibnamefont {Schettler}},\
  and\ \bibinfo {author} {\bibfnamefont {J.}~\bibnamefont
  {Schaffner-Bielich}},\ }\bibfield  {title} {\bibinfo {title} {{The
  Cosmological QCD Phase Transition Revisited}},\ }\href
  {https://doi.org/10.1016/j.ppnp.2011.01.017} {\bibfield  {journal} {\bibinfo
  {journal} {Prog. Part. Nucl. Phys.}\ }\textbf {\bibinfo {volume} {66}},\
  \bibinfo {pages} {266} (\bibinfo {year} {2011})},\ \Eprint
  {https://arxiv.org/abs/1012.3342} {arXiv:1012.3342 [astro-ph.CO]}
  \BibitemShut {NoStop}%
\bibitem [{\citenamefont {Schettler}\ \emph {et~al.}(2011)\citenamefont
  {Schettler}, \citenamefont {Boeckel},\ and\ \citenamefont
  {Schaffner-Bielich}}]{Schettler:2010dp}%
  \BibitemOpen
  \bibfield  {author} {\bibinfo {author} {\bibfnamefont {S.}~\bibnamefont
  {Schettler}}, \bibinfo {author} {\bibfnamefont {T.}~\bibnamefont {Boeckel}},\
  and\ \bibinfo {author} {\bibfnamefont {J.}~\bibnamefont
  {Schaffner-Bielich}},\ }\bibfield  {title} {\bibinfo {title} {{Imprints of
  the QCD Phase Transition on the Spectrum of Gravitational Waves}},\ }\href
  {https://doi.org/10.1103/PhysRevD.83.064030} {\bibfield  {journal} {\bibinfo
  {journal} {Phys. Rev. D}\ }\textbf {\bibinfo {volume} {83}},\ \bibinfo
  {pages} {064030} (\bibinfo {year} {2011})},\ \Eprint
  {https://arxiv.org/abs/1010.4857} {arXiv:1010.4857 [astro-ph.CO]}
  \BibitemShut {NoStop}%
\bibitem [{\citenamefont {Boeckel}\ and\ \citenamefont
  {Schaffner-Bielich}(2012)}]{Boeckel:2011yj}%
  \BibitemOpen
  \bibfield  {author} {\bibinfo {author} {\bibfnamefont {T.}~\bibnamefont
  {Boeckel}}\ and\ \bibinfo {author} {\bibfnamefont {J.}~\bibnamefont
  {Schaffner-Bielich}},\ }\bibfield  {title} {\bibinfo {title} {{A little
  inflation at the cosmological QCD phase transition}},\ }\href
  {https://doi.org/10.1103/PhysRevD.85.103506} {\bibfield  {journal} {\bibinfo
  {journal} {Phys. Rev. D}\ }\textbf {\bibinfo {volume} {85}},\ \bibinfo
  {pages} {103506} (\bibinfo {year} {2012})},\ \Eprint
  {https://arxiv.org/abs/1105.0832} {arXiv:1105.0832 [astro-ph.CO]}
  \BibitemShut {NoStop}%
\bibitem [{\citenamefont {Friedberg}\ and\ \citenamefont
  {Lee}(1978)}]{PhysRevD.18.2623}%
  \BibitemOpen
  \bibfield  {author} {\bibinfo {author} {\bibfnamefont {R.}~\bibnamefont
  {Friedberg}}\ and\ \bibinfo {author} {\bibfnamefont {T.~D.}\ \bibnamefont
  {Lee}},\ }\bibfield  {title} {\bibinfo {title} {Quantum chromodynamics and
  the soliton model of hadrons},\ }\href
  {https://doi.org/10.1103/PhysRevD.18.2623} {\bibfield  {journal} {\bibinfo
  {journal} {Phys. Rev. D}\ }\textbf {\bibinfo {volume} {18}},\ \bibinfo
  {pages} {2623} (\bibinfo {year} {1978})}\BibitemShut {NoStop}%
\bibitem [{\citenamefont {Friedberg}\ and\ \citenamefont
  {Lee}(1977{\natexlab{a}})}]{PhysRevD.16.1096}%
  \BibitemOpen
  \bibfield  {author} {\bibinfo {author} {\bibfnamefont {R.}~\bibnamefont
  {Friedberg}}\ and\ \bibinfo {author} {\bibfnamefont {T.~D.}\ \bibnamefont
  {Lee}},\ }\bibfield  {title} {\bibinfo {title} {Fermion-field nontopological
  solitons. ii. models for hadrons},\ }\href
  {https://doi.org/10.1103/PhysRevD.16.1096} {\bibfield  {journal} {\bibinfo
  {journal} {Phys. Rev. D}\ }\textbf {\bibinfo {volume} {16}},\ \bibinfo
  {pages} {1096} (\bibinfo {year} {1977}{\natexlab{a}})}\BibitemShut {NoStop}%
\bibitem [{\citenamefont {Friedberg}\ and\ \citenamefont
  {Lee}(1977{\natexlab{b}})}]{PhysRevD.15.1694}%
  \BibitemOpen
  \bibfield  {author} {\bibinfo {author} {\bibfnamefont {R.}~\bibnamefont
  {Friedberg}}\ and\ \bibinfo {author} {\bibfnamefont {T.~D.}\ \bibnamefont
  {Lee}},\ }\bibfield  {title} {\bibinfo {title} {Fermion-field nontopological
  solitons},\ }\href {https://doi.org/10.1103/PhysRevD.15.1694} {\bibfield
  {journal} {\bibinfo  {journal} {Phys. Rev. D}\ }\textbf {\bibinfo {volume}
  {15}},\ \bibinfo {pages} {1694} (\bibinfo {year}
  {1977}{\natexlab{b}})}\BibitemShut {NoStop}%
\bibitem [{\citenamefont {Wang}\ \emph
  {et~al.}(2024{\natexlab{a}})\citenamefont {Wang}, \citenamefont {Yu},\ and\
  \citenamefont {Mao}}]{Wang:2023omt}%
  \BibitemOpen
  \bibfield  {author} {\bibinfo {author} {\bibfnamefont {J.}~\bibnamefont
  {Wang}}, \bibinfo {author} {\bibfnamefont {Z.}~\bibnamefont {Yu}},\ and\
  \bibinfo {author} {\bibfnamefont {H.}~\bibnamefont {Mao}},\ }\bibfield
  {title} {\bibinfo {title} {{Bubble nucleation in the two-flavor quark-meson
  model*}},\ }\href {https://doi.org/10.1088/1674-1137/ad2a4b} {\bibfield
  {journal} {\bibinfo  {journal} {Chin. Phys. C}\ }\textbf {\bibinfo {volume}
  {48}},\ \bibinfo {pages} {053105} (\bibinfo {year} {2024}{\natexlab{a}})},\
  \Eprint {https://arxiv.org/abs/2309.13529} {arXiv:2309.13529 [hep-ph]}
  \BibitemShut {NoStop}%
\bibitem [{\citenamefont {Wang}\ \emph
  {et~al.}(2024{\natexlab{b}})\citenamefont {Wang}, \citenamefont {Jin},\ and\
  \citenamefont {Mao}}]{Wang:2023pmn}%
  \BibitemOpen
  \bibfield  {author} {\bibinfo {author} {\bibfnamefont {J.}~\bibnamefont
  {Wang}}, \bibinfo {author} {\bibfnamefont {J.}~\bibnamefont {Jin}},\ and\
  \bibinfo {author} {\bibfnamefont {H.}~\bibnamefont {Mao}},\ }\bibfield
  {title} {\bibinfo {title} {{Bubble dynamics in the Polyakov quark-meson
  model}},\ }\href {https://doi.org/10.3390/sym16070893} {\bibfield  {journal}
  {\bibinfo  {journal} {Symmetry}\ }\textbf {\bibinfo {volume} {16}},\ \bibinfo
  {pages} {893} (\bibinfo {year} {2024}{\natexlab{b}})},\ \Eprint
  {https://arxiv.org/abs/2311.13175} {arXiv:2311.13175 [hep-ph]} \BibitemShut
  {NoStop}%
\bibitem [{\citenamefont {Gupta}\ and\ \citenamefont
  {Tiwari}(2012)}]{Gupta:2011ez}%
  \BibitemOpen
  \bibfield  {author} {\bibinfo {author} {\bibfnamefont {U.~S.}\ \bibnamefont
  {Gupta}}\ and\ \bibinfo {author} {\bibfnamefont {V.~K.}\ \bibnamefont
  {Tiwari}},\ }\bibfield  {title} {\bibinfo {title} {{Revisiting the Phase
  Structure of the Polyakov-quark-meson Model in the presence of Vacuum Fermion
  Fluctuation}},\ }\href {https://doi.org/10.1103/PhysRevD.85.014010}
  {\bibfield  {journal} {\bibinfo  {journal} {Phys. Rev. D}\ }\textbf {\bibinfo
  {volume} {85}},\ \bibinfo {pages} {014010} (\bibinfo {year} {2012})},\
  \Eprint {https://arxiv.org/abs/1107.1312} {arXiv:1107.1312 [hep-ph]}
  \BibitemShut {NoStop}%
\bibitem [{\citenamefont {Shao}\ \emph {et~al.}(2025)\citenamefont {Shao},
  \citenamefont {Mao},\ and\ \citenamefont {Huang}}]{Shao:2024dxt}%
  \BibitemOpen
  \bibfield  {author} {\bibinfo {author} {\bibfnamefont {J.}~\bibnamefont
  {Shao}}, \bibinfo {author} {\bibfnamefont {H.}~\bibnamefont {Mao}},\ and\
  \bibinfo {author} {\bibfnamefont {M.}~\bibnamefont {Huang}},\ }\bibfield
  {title} {\bibinfo {title} {{Transition rate and gravitational wave spectrum
  from first-order QCD phase transitions}},\ }\href
  {https://doi.org/10.1103/PhysRevD.111.023052} {\bibfield  {journal} {\bibinfo
   {journal} {Phys. Rev. D}\ }\textbf {\bibinfo {volume} {111}},\ \bibinfo
  {pages} {023052} (\bibinfo {year} {2025})},\ \Eprint
  {https://arxiv.org/abs/2410.06780} {arXiv:2410.06780 [hep-ph]} \BibitemShut
  {NoStop}%
\bibitem [{\citenamefont {Chen}\ \emph {et~al.}(2023)\citenamefont {Chen},
  \citenamefont {Li},\ and\ \citenamefont {Huang}}]{Chen:2022cgj}%
  \BibitemOpen
  \bibfield  {author} {\bibinfo {author} {\bibfnamefont {Y.}~\bibnamefont
  {Chen}}, \bibinfo {author} {\bibfnamefont {D.}~\bibnamefont {Li}},\ and\
  \bibinfo {author} {\bibfnamefont {M.}~\bibnamefont {Huang}},\ }\bibfield
  {title} {\bibinfo {title} {{Bubble nucleation and gravitational waves from
  holography in the probe approximation}},\ }\href
  {https://doi.org/10.1007/JHEP07(2023)225} {\bibfield  {journal} {\bibinfo
  {journal} {JHEP}\ }\textbf {\bibinfo {volume} {07}},\ \bibinfo {pages}
  {225}},\ \Eprint {https://arxiv.org/abs/2212.06591} {arXiv:2212.06591
  [hep-ph]} \BibitemShut {NoStop}%
\bibitem [{\citenamefont {Morgante}\ \emph {et~al.}(2023)\citenamefont
  {Morgante}, \citenamefont {Ramberg},\ and\ \citenamefont
  {Schwaller}}]{Morgante:2022zvc}%
  \BibitemOpen
  \bibfield  {author} {\bibinfo {author} {\bibfnamefont {E.}~\bibnamefont
  {Morgante}}, \bibinfo {author} {\bibfnamefont {N.}~\bibnamefont {Ramberg}},\
  and\ \bibinfo {author} {\bibfnamefont {P.}~\bibnamefont {Schwaller}},\
  }\bibfield  {title} {\bibinfo {title} {{Gravitational waves from dark SU(3)
  Yang-Mills theory}},\ }\href {https://doi.org/10.1103/PhysRevD.107.036010}
  {\bibfield  {journal} {\bibinfo  {journal} {Phys. Rev. D}\ }\textbf {\bibinfo
  {volume} {107}},\ \bibinfo {pages} {036010} (\bibinfo {year} {2023})},\
  \Eprint {https://arxiv.org/abs/2210.11821} {arXiv:2210.11821 [hep-ph]}
  \BibitemShut {NoStop}%
\bibitem [{\citenamefont {Helmboldt}\ \emph {et~al.}(2019)\citenamefont
  {Helmboldt}, \citenamefont {Kubo},\ and\ \citenamefont {van~der
  Woude}}]{Helmboldt:2019pan}%
  \BibitemOpen
  \bibfield  {author} {\bibinfo {author} {\bibfnamefont {A.~J.}\ \bibnamefont
  {Helmboldt}}, \bibinfo {author} {\bibfnamefont {J.}~\bibnamefont {Kubo}},\
  and\ \bibinfo {author} {\bibfnamefont {S.}~\bibnamefont {van~der Woude}},\
  }\bibfield  {title} {\bibinfo {title} {{Observational prospects for
  gravitational waves from hidden or dark chiral phase transitions}},\ }\href
  {https://doi.org/10.1103/PhysRevD.100.055025} {\bibfield  {journal} {\bibinfo
   {journal} {Phys. Rev. D}\ }\textbf {\bibinfo {volume} {100}},\ \bibinfo
  {pages} {055025} (\bibinfo {year} {2019})},\ \Eprint
  {https://arxiv.org/abs/1904.07891} {arXiv:1904.07891 [hep-ph]} \BibitemShut
  {NoStop}%
\bibitem [{\citenamefont {Detar}\ and\ \citenamefont
  {Kunihiro}(1989)}]{Detar:1988kn}%
  \BibitemOpen
  \bibfield  {author} {\bibinfo {author} {\bibfnamefont {C.~E.}\ \bibnamefont
  {Detar}}\ and\ \bibinfo {author} {\bibfnamefont {T.}~\bibnamefont
  {Kunihiro}},\ }\bibfield  {title} {\bibinfo {title} {{Linear $\sigma$ Model
  With Parity Doubling}},\ }\href {https://doi.org/10.1103/PhysRevD.39.2805}
  {\bibfield  {journal} {\bibinfo  {journal} {Phys. Rev. D}\ }\textbf {\bibinfo
  {volume} {39}},\ \bibinfo {pages} {2805} (\bibinfo {year}
  {1989})}\BibitemShut {NoStop}%
\bibitem [{\citenamefont {Jido}\ \emph {et~al.}(2001)\citenamefont {Jido},
  \citenamefont {Oka},\ and\ \citenamefont {Hosaka}}]{Jido:2001nt}%
  \BibitemOpen
  \bibfield  {author} {\bibinfo {author} {\bibfnamefont {D.}~\bibnamefont
  {Jido}}, \bibinfo {author} {\bibfnamefont {M.}~\bibnamefont {Oka}},\ and\
  \bibinfo {author} {\bibfnamefont {A.}~\bibnamefont {Hosaka}},\ }\bibfield
  {title} {\bibinfo {title} {{Chiral symmetry of baryons}},\ }\href
  {https://doi.org/10.1143/PTP.106.873} {\bibfield  {journal} {\bibinfo
  {journal} {Prog. Theor. Phys.}\ }\textbf {\bibinfo {volume} {106}},\ \bibinfo
  {pages} {873} (\bibinfo {year} {2001})},\ \Eprint
  {https://arxiv.org/abs/hep-ph/0110005} {arXiv:hep-ph/0110005} \BibitemShut
  {NoStop}%
\bibitem [{\citenamefont {Gallas}\ \emph {et~al.}(2010)\citenamefont {Gallas},
  \citenamefont {Giacosa},\ and\ \citenamefont {Rischke}}]{Gallas:2009qp}%
  \BibitemOpen
  \bibfield  {author} {\bibinfo {author} {\bibfnamefont {S.}~\bibnamefont
  {Gallas}}, \bibinfo {author} {\bibfnamefont {F.}~\bibnamefont {Giacosa}},\
  and\ \bibinfo {author} {\bibfnamefont {D.~H.}\ \bibnamefont {Rischke}},\
  }\bibfield  {title} {\bibinfo {title} {{Vacuum phenomenology of the chiral
  partner of the nucleon in a linear sigma model with vector mesons}},\ }\href
  {https://doi.org/10.1103/PhysRevD.82.014004} {\bibfield  {journal} {\bibinfo
  {journal} {Phys. Rev. D}\ }\textbf {\bibinfo {volume} {82}},\ \bibinfo
  {pages} {014004} (\bibinfo {year} {2010})},\ \Eprint
  {https://arxiv.org/abs/0907.5084} {arXiv:0907.5084 [hep-ph]} \BibitemShut
  {NoStop}%
\bibitem [{\citenamefont {Steinheimer}\ \emph {et~al.}(2011)\citenamefont
  {Steinheimer}, \citenamefont {Schramm},\ and\ \citenamefont
  {Stocker}}]{Steinheimer:2011ea}%
  \BibitemOpen
  \bibfield  {author} {\bibinfo {author} {\bibfnamefont {J.}~\bibnamefont
  {Steinheimer}}, \bibinfo {author} {\bibfnamefont {S.}~\bibnamefont
  {Schramm}},\ and\ \bibinfo {author} {\bibfnamefont {H.}~\bibnamefont
  {Stocker}},\ }\bibfield  {title} {\bibinfo {title} {{The hadronic SU(3)
  Parity Doublet Model for Dense Matter, its extension to quarks and the
  strange equation of state}},\ }\href
  {https://doi.org/10.1103/PhysRevC.84.045208} {\bibfield  {journal} {\bibinfo
  {journal} {Phys. Rev. C}\ }\textbf {\bibinfo {volume} {84}},\ \bibinfo
  {pages} {045208} (\bibinfo {year} {2011})},\ \Eprint
  {https://arxiv.org/abs/1108.2596} {arXiv:1108.2596 [hep-ph]} \BibitemShut
  {NoStop}%
\bibitem [{\citenamefont {Zschiesche}\ \emph {et~al.}(2007)\citenamefont
  {Zschiesche}, \citenamefont {Tolos}, \citenamefont {Schaffner-Bielich},\ and\
  \citenamefont {Pisarski}}]{Zschiesche:2006zj}%
  \BibitemOpen
  \bibfield  {author} {\bibinfo {author} {\bibfnamefont {D.}~\bibnamefont
  {Zschiesche}}, \bibinfo {author} {\bibfnamefont {L.}~\bibnamefont {Tolos}},
  \bibinfo {author} {\bibfnamefont {J.}~\bibnamefont {Schaffner-Bielich}},\
  and\ \bibinfo {author} {\bibfnamefont {R.~D.}\ \bibnamefont {Pisarski}},\
  }\bibfield  {title} {\bibinfo {title} {{Cold, dense nuclear matter in a SU(2)
  parity doublet model}},\ }\href {https://doi.org/10.1103/PhysRevC.75.055202}
  {\bibfield  {journal} {\bibinfo  {journal} {Phys. Rev. C}\ }\textbf {\bibinfo
  {volume} {75}},\ \bibinfo {pages} {055202} (\bibinfo {year} {2007})},\
  \Eprint {https://arxiv.org/abs/nucl-th/0608044} {arXiv:nucl-th/0608044}
  \BibitemShut {NoStop}%
\bibitem [{\citenamefont {Dexheimer}\ \emph {et~al.}(2008)\citenamefont
  {Dexheimer}, \citenamefont {Schramm},\ and\ \citenamefont
  {Zschiesche}}]{Dexheimer:2007tn}%
  \BibitemOpen
  \bibfield  {author} {\bibinfo {author} {\bibfnamefont {V.}~\bibnamefont
  {Dexheimer}}, \bibinfo {author} {\bibfnamefont {S.}~\bibnamefont {Schramm}},\
  and\ \bibinfo {author} {\bibfnamefont {D.}~\bibnamefont {Zschiesche}},\
  }\bibfield  {title} {\bibinfo {title} {{Nuclear matter and neutron stars in a
  parity doublet model}},\ }\href {https://doi.org/10.1103/PhysRevC.77.025803}
  {\bibfield  {journal} {\bibinfo  {journal} {Phys. Rev. C}\ }\textbf {\bibinfo
  {volume} {77}},\ \bibinfo {pages} {025803} (\bibinfo {year} {2008})},\
  \Eprint {https://arxiv.org/abs/0710.4192} {arXiv:0710.4192 [nucl-th]}
  \BibitemShut {NoStop}%
\bibitem [{\citenamefont {Sasaki}\ \emph {et~al.}(2011)\citenamefont {Sasaki},
  \citenamefont {Lee}, \citenamefont {Paeng},\ and\ \citenamefont
  {Rho}}]{Sasaki:2011ff}%
  \BibitemOpen
  \bibfield  {author} {\bibinfo {author} {\bibfnamefont {C.}~\bibnamefont
  {Sasaki}}, \bibinfo {author} {\bibfnamefont {H.~K.}\ \bibnamefont {Lee}},
  \bibinfo {author} {\bibfnamefont {W.-G.}\ \bibnamefont {Paeng}},\ and\
  \bibinfo {author} {\bibfnamefont {M.}~\bibnamefont {Rho}},\ }\bibfield
  {title} {\bibinfo {title} {{Conformal anomaly and the vector coupling in
  dense matter}},\ }\href {https://doi.org/10.1103/PhysRevD.84.034011}
  {\bibfield  {journal} {\bibinfo  {journal} {Phys. Rev. D}\ }\textbf {\bibinfo
  {volume} {84}},\ \bibinfo {pages} {034011} (\bibinfo {year} {2011})},\
  \Eprint {https://arxiv.org/abs/1103.0184} {arXiv:1103.0184 [hep-ph]}
  \BibitemShut {NoStop}%
\bibitem [{\citenamefont {Motohiro}\ \emph {et~al.}(2015)\citenamefont
  {Motohiro}, \citenamefont {Kim},\ and\ \citenamefont
  {Harada}}]{Motohiro:2015taa}%
  \BibitemOpen
  \bibfield  {author} {\bibinfo {author} {\bibfnamefont {Y.}~\bibnamefont
  {Motohiro}}, \bibinfo {author} {\bibfnamefont {Y.}~\bibnamefont {Kim}},\ and\
  \bibinfo {author} {\bibfnamefont {M.}~\bibnamefont {Harada}},\ }\bibfield
  {title} {\bibinfo {title} {{Asymmetric nuclear matter in a parity doublet
  model with hidden local symmetry}},\ }\href
  {https://doi.org/10.1103/PhysRevC.92.025201} {\bibfield  {journal} {\bibinfo
  {journal} {Phys. Rev. C}\ }\textbf {\bibinfo {volume} {92}},\ \bibinfo
  {pages} {025201} (\bibinfo {year} {2015})},\ \bibinfo {note} {[Erratum:
  Phys.Rev.C 95, 059903 (2017)]},\ \Eprint {https://arxiv.org/abs/1505.00988}
  {arXiv:1505.00988 [nucl-th]} \BibitemShut {NoStop}%
\bibitem [{\citenamefont {Gao}\ and\ \citenamefont
  {Hosaka}(2026)}]{Gao:2025eax}%
  \BibitemOpen
  \bibfield  {author} {\bibinfo {author} {\bibfnamefont {B.}~\bibnamefont
  {Gao}}\ and\ \bibinfo {author} {\bibfnamefont {A.}~\bibnamefont {Hosaka}},\
  }\bibfield  {title} {\bibinfo {title} {{Linear realization of an SU(3) parity
  doublet model for octet baryons with a bad diquark}},\ }\href
  {https://doi.org/10.1103/s39n-mlw2} {\bibfield  {journal} {\bibinfo
  {journal} {Phys. Rev. D}\ }\textbf {\bibinfo {volume} {113}},\ \bibinfo
  {pages} {036008} (\bibinfo {year} {2026})},\ \Eprint
  {https://arxiv.org/abs/2512.01192} {arXiv:2512.01192 [hep-ph]} \BibitemShut
  {NoStop}%
\bibitem [{\citenamefont {Marczenko}\ and\ \citenamefont
  {Sasaki}(2018)}]{Marczenko:2017huu}%
  \BibitemOpen
  \bibfield  {author} {\bibinfo {author} {\bibfnamefont {M.}~\bibnamefont
  {Marczenko}}\ and\ \bibinfo {author} {\bibfnamefont {C.}~\bibnamefont
  {Sasaki}},\ }\bibfield  {title} {\bibinfo {title} {{Net-baryon number
  fluctuations in the Hybrid Quark-Meson-Nucleon model at finite density}},\
  }\href {https://doi.org/10.1103/PhysRevD.97.036011} {\bibfield  {journal}
  {\bibinfo  {journal} {Phys. Rev. D}\ }\textbf {\bibinfo {volume} {97}},\
  \bibinfo {pages} {036011} (\bibinfo {year} {2018})},\ \Eprint
  {https://arxiv.org/abs/1711.05521} {arXiv:1711.05521 [hep-ph]} \BibitemShut
  {NoStop}%
\bibitem [{\citenamefont {Minamikawa}\ \emph {et~al.}(2021)\citenamefont
  {Minamikawa}, \citenamefont {Kojo},\ and\ \citenamefont
  {Harada}}]{Minamikawa:2020jfj}%
  \BibitemOpen
  \bibfield  {author} {\bibinfo {author} {\bibfnamefont {T.}~\bibnamefont
  {Minamikawa}}, \bibinfo {author} {\bibfnamefont {T.}~\bibnamefont {Kojo}},\
  and\ \bibinfo {author} {\bibfnamefont {M.}~\bibnamefont {Harada}},\
  }\bibfield  {title} {\bibinfo {title} {{Quark-hadron crossover equations of
  state for neutron stars: constraining the chiral invariant mass in a parity
  doublet model}},\ }\href {https://doi.org/10.1103/PhysRevC.103.045205}
  {\bibfield  {journal} {\bibinfo  {journal} {Phys. Rev. C}\ }\textbf {\bibinfo
  {volume} {103}},\ \bibinfo {pages} {045205} (\bibinfo {year} {2021})},\
  \Eprint {https://arxiv.org/abs/2011.13684} {arXiv:2011.13684 [nucl-th]}
  \BibitemShut {NoStop}%
\bibitem [{\citenamefont {Marczenko}\ \emph {et~al.}(2022)\citenamefont
  {Marczenko}, \citenamefont {Redlich},\ and\ \citenamefont
  {Sasaki}}]{Marczenko:2021uaj}%
  \BibitemOpen
  \bibfield  {author} {\bibinfo {author} {\bibfnamefont {M.}~\bibnamefont
  {Marczenko}}, \bibinfo {author} {\bibfnamefont {K.}~\bibnamefont {Redlich}},\
  and\ \bibinfo {author} {\bibfnamefont {C.}~\bibnamefont {Sasaki}},\
  }\bibfield  {title} {\bibinfo {title} {{Reconciling Multi-messenger
  Constraints with Chiral Symmetry Restoration}},\ }\href
  {https://doi.org/10.3847/2041-8213/ac4b61} {\bibfield  {journal} {\bibinfo
  {journal} {Astrophys. J. Lett.}\ }\textbf {\bibinfo {volume} {925}},\
  \bibinfo {pages} {L23} (\bibinfo {year} {2022})},\ \Eprint
  {https://arxiv.org/abs/2110.11056} {arXiv:2110.11056 [nucl-th]} \BibitemShut
  {NoStop}%
\bibitem [{\citenamefont {Gao}\ \emph {et~al.}(2024)\citenamefont {Gao},
  \citenamefont {Yan},\ and\ \citenamefont {Harada}}]{Gao:2024chh}%
  \BibitemOpen
  \bibfield  {author} {\bibinfo {author} {\bibfnamefont {B.}~\bibnamefont
  {Gao}}, \bibinfo {author} {\bibfnamefont {Y.}~\bibnamefont {Yan}},\ and\
  \bibinfo {author} {\bibfnamefont {M.}~\bibnamefont {Harada}},\ }\bibfield
  {title} {\bibinfo {title} {{Reconciling constraints from the supernova
  remnant HESS J1731-347 with the parity doublet model}},\ }\href
  {https://doi.org/10.1103/PhysRevC.109.065807} {\bibfield  {journal} {\bibinfo
   {journal} {Phys. Rev. C}\ }\textbf {\bibinfo {volume} {109}},\ \bibinfo
  {pages} {065807} (\bibinfo {year} {2024})},\ \Eprint
  {https://arxiv.org/abs/2404.04786} {arXiv:2404.04786 [nucl-th]} \BibitemShut
  {NoStop}%
\bibitem [{\citenamefont {Gao}\ \emph {et~al.}(2026)\citenamefont {Gao},
  \citenamefont {Liu}, \citenamefont {Harada},\ and\ \citenamefont
  {Ma}}]{Gao:2025nkg}%
  \BibitemOpen
  \bibfield  {author} {\bibinfo {author} {\bibfnamefont {B.}~\bibnamefont
  {Gao}}, \bibinfo {author} {\bibfnamefont {X.}~\bibnamefont {Liu}}, \bibinfo
  {author} {\bibfnamefont {M.}~\bibnamefont {Harada}},\ and\ \bibinfo {author}
  {\bibfnamefont {Y.-L.}\ \bibnamefont {Ma}},\ }\bibfield  {title} {\bibinfo
  {title} {{Implication of neutron star observations to the origin of nucleon
  mass}},\ }\href {https://doi.org/10.1007/s11433-025-2839-7} {\bibfield
  {journal} {\bibinfo  {journal} {Sci. China Phys. Mech. Astron.}\ }\textbf
  {\bibinfo {volume} {69}},\ \bibinfo {pages} {232011} (\bibinfo {year}
  {2026})},\ \Eprint {https://arxiv.org/abs/2508.00243} {arXiv:2508.00243
  [nucl-th]} \BibitemShut {NoStop}%
\bibitem [{\citenamefont {Gao}\ \emph {et~al.}(2025)\citenamefont {Gao},
  \citenamefont {Kong},\ and\ \citenamefont {Ma}}]{Gao:2025vdc}%
  \BibitemOpen
  \bibfield  {author} {\bibinfo {author} {\bibfnamefont {B.}~\bibnamefont
  {Gao}}, \bibinfo {author} {\bibfnamefont {Y.-K.}\ \bibnamefont {Kong}},\ and\
  \bibinfo {author} {\bibfnamefont {Y.-L.}\ \bibnamefont {Ma}},\ }\bibfield
  {title} {\bibinfo {title} {{Origin of nucleon mass in the light of PSR
  J0614-3329 with quark-hadron crossover}},\ }\href
  {https://doi.org/10.1103/xwkv-s8lg} {\bibfield  {journal} {\bibinfo
  {journal} {Phys. Rev. D}\ }\textbf {\bibinfo {volume} {112}},\ \bibinfo
  {pages} {083041} (\bibinfo {year} {2025})},\ \Eprint
  {https://arxiv.org/abs/2509.03008} {arXiv:2509.03008 [nucl-th]} \BibitemShut
  {NoStop}%
\bibitem [{\citenamefont {Gao}\ and\ \citenamefont
  {Marczenko}(2026)}]{Gao:2025okn}%
  \BibitemOpen
  \bibfield  {author} {\bibinfo {author} {\bibfnamefont {B.}~\bibnamefont
  {Gao}}\ and\ \bibinfo {author} {\bibfnamefont {M.}~\bibnamefont
  {Marczenko}},\ }\bibfield  {title} {\bibinfo {title} {{Suppression of a
  dynamical momentum-space shell by chiral symmetry}},\ }\href
  {https://doi.org/10.1103/zlsk-vzmx} {\bibfield  {journal} {\bibinfo
  {journal} {Phys. Rev. C}\ }\textbf {\bibinfo {volume} {113}},\ \bibinfo
  {pages} {015205} (\bibinfo {year} {2026})},\ \Eprint
  {https://arxiv.org/abs/2509.03138} {arXiv:2509.03138 [nucl-th]} \BibitemShut
  {NoStop}%
\bibitem [{\citenamefont {Yuan}\ \emph {et~al.}(2025)\citenamefont {Yuan},
  \citenamefont {Gao}, \citenamefont {Yan},\ and\ \citenamefont
  {Xu}}]{Yuan:2025dft}%
  \BibitemOpen
  \bibfield  {author} {\bibinfo {author} {\bibfnamefont {W.-L.}\ \bibnamefont
  {Yuan}}, \bibinfo {author} {\bibfnamefont {B.}~\bibnamefont {Gao}}, \bibinfo
  {author} {\bibfnamefont {Y.}~\bibnamefont {Yan}},\ and\ \bibinfo {author}
  {\bibfnamefont {R.}~\bibnamefont {Xu}},\ }\bibfield  {title} {\bibinfo
  {title} {{Hybrid stars with large quark cores within the parity doublet model
  and modified NJL model}},\ }\href {https://doi.org/10.1103/nlhh-xqjp}
  {\bibfield  {journal} {\bibinfo  {journal} {Phys. Rev. D}\ }\textbf {\bibinfo
  {volume} {112}},\ \bibinfo {pages} {023019} (\bibinfo {year} {2025})},\
  \Eprint {https://arxiv.org/abs/2502.17859} {arXiv:2502.17859 [nucl-th]}
  \BibitemShut {NoStop}%
\bibitem [{\citenamefont {Fraga}\ \emph {et~al.}(2023)\citenamefont {Fraga},
  \citenamefont {da~Mata},\ and\ \citenamefont
  {Schaffner-Bielich}}]{Fraga:2023wtd}%
  \BibitemOpen
  \bibfield  {author} {\bibinfo {author} {\bibfnamefont {E.~S.}\ \bibnamefont
  {Fraga}}, \bibinfo {author} {\bibfnamefont {R.}~\bibnamefont {da~Mata}},\
  and\ \bibinfo {author} {\bibfnamefont {J.}~\bibnamefont
  {Schaffner-Bielich}},\ }\bibfield  {title} {\bibinfo {title} {{SU(3) parity
  doubling in cold neutron star matter}},\ }\href
  {https://doi.org/10.1103/PhysRevD.108.116003} {\bibfield  {journal} {\bibinfo
   {journal} {Phys. Rev. D}\ }\textbf {\bibinfo {volume} {108}},\ \bibinfo
  {pages} {116003} (\bibinfo {year} {2023})},\ \Eprint
  {https://arxiv.org/abs/2309.02368} {arXiv:2309.02368 [hep-ph]} \BibitemShut
  {NoStop}%
\bibitem [{\citenamefont {Gao}(2026)}]{Gao:2026scv}%
  \BibitemOpen
  \bibfield  {author} {\bibinfo {author} {\bibfnamefont {B.}~\bibnamefont
  {Gao}},\ }\bibfield  {title} {\bibinfo {title} {{Chiral symmetry restoration
  and hyperon suppression in neutron stars}},\ }\href@noop {} {\  (\bibinfo
  {year} {2026})},\ \Eprint {https://arxiv.org/abs/2602.12503}
  {arXiv:2602.12503 [nucl-th]} \BibitemShut {NoStop}%
\bibitem [{\citenamefont {Marczenko}(2025)}]{Marczenko:2025kpv}%
  \BibitemOpen
  \bibfield  {author} {\bibinfo {author} {\bibfnamefont {M.}~\bibnamefont
  {Marczenko}},\ }\bibfield  {title} {\bibinfo {title} {{Proton-neutron
  correlations in baryon-number fluctuations near the liquid-gas transition}}\
  }(\bibinfo {year} {2025})\ \Eprint {https://arxiv.org/abs/2512.20258}
  {arXiv:2512.20258 [nucl-th]} \BibitemShut {NoStop}%
\bibitem [{\citenamefont {Kong}\ \emph {et~al.}(2025)\citenamefont {Kong},
  \citenamefont {Gao},\ and\ \citenamefont {Harada}}]{Kong:2025dwl}%
  \BibitemOpen
  \bibfield  {author} {\bibinfo {author} {\bibfnamefont {Y.-K.}\ \bibnamefont
  {Kong}}, \bibinfo {author} {\bibfnamefont {B.}~\bibnamefont {Gao}},\ and\
  \bibinfo {author} {\bibfnamefont {M.}~\bibnamefont {Harada}},\ }\bibfield
  {title} {\bibinfo {title} {{Chiral Invariant Mass Constraints from HESS
  J1731{\textendash}347 in an Extended Parity Doublet Model with Isovector
  Scalar Meson}},\ }\href {https://doi.org/10.3390/universe11100345} {\bibfield
   {journal} {\bibinfo  {journal} {Universe}\ }\textbf {\bibinfo {volume}
  {11}},\ \bibinfo {pages} {345} (\bibinfo {year} {2025})},\ \Eprint
  {https://arxiv.org/abs/2506.16684} {arXiv:2506.16684 [nucl-th]} \BibitemShut
  {NoStop}%
\bibitem [{\citenamefont {Coleman}(1977)}]{Coleman:1977py}%
  \BibitemOpen
  \bibfield  {author} {\bibinfo {author} {\bibfnamefont {S.~R.}\ \bibnamefont
  {Coleman}},\ }\bibfield  {title} {\bibinfo {title} {{The Fate of the False
  Vacuum. 1. Semiclassical Theory}},\ }\href
  {https://doi.org/10.1103/PhysRevD.16.1248} {\bibfield  {journal} {\bibinfo
  {journal} {Phys. Rev. D}\ }\textbf {\bibinfo {volume} {15}},\ \bibinfo
  {pages} {2929} (\bibinfo {year} {1977})},\ \bibinfo {note} {[Erratum:
  Phys.Rev.D 16, 1248 (1977)]}\BibitemShut {NoStop}%
\bibitem [{\citenamefont {Ellis}\ \emph
  {et~al.}(2020{\natexlab{a}})\citenamefont {Ellis}, \citenamefont {Lewicki},\
  and\ \citenamefont {No}}]{Ellis:2020awk}%
  \BibitemOpen
  \bibfield  {author} {\bibinfo {author} {\bibfnamefont {J.}~\bibnamefont
  {Ellis}}, \bibinfo {author} {\bibfnamefont {M.}~\bibnamefont {Lewicki}},\
  and\ \bibinfo {author} {\bibfnamefont {J.~M.}\ \bibnamefont {No}},\
  }\bibfield  {title} {\bibinfo {title} {{Gravitational waves from first-order
  cosmological phase transitions: lifetime of the sound wave source}},\ }\href
  {https://doi.org/10.1088/1475-7516/2020/07/050} {\bibfield  {journal}
  {\bibinfo  {journal} {JCAP}\ }\textbf {\bibinfo {volume} {07}},\ \bibinfo
  {pages} {050}},\ \Eprint {https://arxiv.org/abs/2003.07360} {arXiv:2003.07360
  [hep-ph]} \BibitemShut {NoStop}%
\bibitem [{\citenamefont {Binetruy}\ \emph {et~al.}(2012)\citenamefont
  {Binetruy}, \citenamefont {Bohe}, \citenamefont {Caprini},\ and\
  \citenamefont {Dufaux}}]{Binetruy:2012ze}%
  \BibitemOpen
  \bibfield  {author} {\bibinfo {author} {\bibfnamefont {P.}~\bibnamefont
  {Binetruy}}, \bibinfo {author} {\bibfnamefont {A.}~\bibnamefont {Bohe}},
  \bibinfo {author} {\bibfnamefont {C.}~\bibnamefont {Caprini}},\ and\ \bibinfo
  {author} {\bibfnamefont {J.-F.}\ \bibnamefont {Dufaux}},\ }\bibfield  {title}
  {\bibinfo {title} {{Cosmological Backgrounds of Gravitational Waves and
  eLISA/NGO: Phase Transitions, Cosmic Strings and Other Sources}},\ }\href
  {https://doi.org/10.1088/1475-7516/2012/06/027} {\bibfield  {journal}
  {\bibinfo  {journal} {JCAP}\ }\textbf {\bibinfo {volume} {06}},\ \bibinfo
  {pages} {027}},\ \Eprint {https://arxiv.org/abs/1201.0983} {arXiv:1201.0983
  [gr-qc]} \BibitemShut {NoStop}%
\bibitem [{\citenamefont {Eichhorn}\ \emph
  {et~al.}(2021{\natexlab{a}})\citenamefont {Eichhorn}, \citenamefont {Lumma},
  \citenamefont {Pawlowski}, \citenamefont {Reichert},\ and\ \citenamefont
  {Yamada}}]{Eichhorn:2020upj}%
  \BibitemOpen
  \bibfield  {author} {\bibinfo {author} {\bibfnamefont {A.}~\bibnamefont
  {Eichhorn}}, \bibinfo {author} {\bibfnamefont {J.}~\bibnamefont {Lumma}},
  \bibinfo {author} {\bibfnamefont {J.~M.}\ \bibnamefont {Pawlowski}}, \bibinfo
  {author} {\bibfnamefont {M.}~\bibnamefont {Reichert}},\ and\ \bibinfo
  {author} {\bibfnamefont {M.}~\bibnamefont {Yamada}},\ }\bibfield  {title}
  {\bibinfo {title} {{Universal gravitational-wave signatures from heavy new
  physics in the electroweak sector}},\ }\href
  {https://doi.org/10.1088/1475-7516/2021/05/006} {\bibfield  {journal}
  {\bibinfo  {journal} {JCAP}\ }\textbf {\bibinfo {volume} {05}},\ \bibinfo
  {pages} {006}},\ \Eprint {https://arxiv.org/abs/2010.00017} {arXiv:2010.00017
  [hep-ph]} \BibitemShut {NoStop}%
\bibitem [{\citenamefont {Ellis}\ \emph
  {et~al.}(2020{\natexlab{b}})\citenamefont {Ellis}, \citenamefont {Lewicki},\
  and\ \citenamefont {No}}]{ellis2020gravitational}%
  \BibitemOpen
  \bibfield  {author} {\bibinfo {author} {\bibfnamefont {J.}~\bibnamefont
  {Ellis}}, \bibinfo {author} {\bibfnamefont {M.}~\bibnamefont {Lewicki}},\
  and\ \bibinfo {author} {\bibfnamefont {J.~M.}\ \bibnamefont {No}},\
  }\bibfield  {title} {\bibinfo {title} {Gravitational waves from first-order
  cosmological phase transitions: lifetime of the sound wave source},\
  }\href@noop {} {\bibfield  {journal} {\bibinfo  {journal} {Journal of
  Cosmology and Astroparticle Physics}\ }\textbf {\bibinfo {volume}
  {2020}}\bibinfo  {number} { (07)},\ \bibinfo {pages} {050}}\BibitemShut
  {NoStop}%
\bibitem [{\citenamefont {Binétruy}\ \emph {et~al.}(2012)\citenamefont
  {Binétruy}, \citenamefont {Bohé}, \citenamefont {Caprini},\ and\
  \citenamefont {Dufaux}}]{RN20}%
  \BibitemOpen
\bibfield  {number} {  }\bibfield  {author} {\bibinfo {author} {\bibfnamefont
  {P.}~\bibnamefont {Binétruy}}, \bibinfo {author} {\bibfnamefont
  {A.}~\bibnamefont {Bohé}}, \bibinfo {author} {\bibfnamefont
  {C.}~\bibnamefont {Caprini}},\ and\ \bibinfo {author} {\bibfnamefont {J.-F.}\
  \bibnamefont {Dufaux}},\ }\bibfield  {title} {\bibinfo {title} {Cosmological
  backgrounds of gravitational waves and elisa/ngo: phase transitions, cosmic
  strings and other sources},\ }\href
  {https://doi.org/10.1088/1475-7516/2012/06/027} {\bibfield  {journal}
  {\bibinfo  {journal} {Journal of Cosmology and Astroparticle Physics}\
  }\textbf {\bibinfo {volume} {2012}},\ \bibinfo {pages} {027}}\BibitemShut
  {NoStop}%
\bibitem [{\citenamefont {Wang}\ \emph
  {et~al.}(2024{\natexlab{c}})\citenamefont {Wang}, \citenamefont {Yan},\ and\
  \citenamefont {Huang}}]{Wang:2024wcs}%
  \BibitemOpen
  \bibfield  {author} {\bibinfo {author} {\bibfnamefont {D.-W.}\ \bibnamefont
  {Wang}}, \bibinfo {author} {\bibfnamefont {Q.-S.}\ \bibnamefont {Yan}},\ and\
  \bibinfo {author} {\bibfnamefont {M.}~\bibnamefont {Huang}},\ }\bibfield
  {title} {\bibinfo {title} {{Bubble wall velocity and gravitational wave in
  the minimal left-right symmetric model}},\ }\href
  {https://doi.org/10.1103/PhysRevD.110.076011} {\bibfield  {journal} {\bibinfo
   {journal} {Phys. Rev. D}\ }\textbf {\bibinfo {volume} {110}},\ \bibinfo
  {pages} {076011} (\bibinfo {year} {2024}{\natexlab{c}})},\ \Eprint
  {https://arxiv.org/abs/2405.01949} {arXiv:2405.01949 [gr-qc]} \BibitemShut
  {NoStop}%
\bibitem [{\citenamefont {Kurki-Suonio}\ and\ \citenamefont
  {Laine}(1996)}]{kurki1996bubble}%
  \BibitemOpen
  \bibfield  {author} {\bibinfo {author} {\bibfnamefont {H.}~\bibnamefont
  {Kurki-Suonio}}\ and\ \bibinfo {author} {\bibfnamefont {M.}~\bibnamefont
  {Laine}},\ }\bibfield  {title} {\bibinfo {title} {Bubble growth and droplet
  decay in cosmological phase transitions},\ }\href@noop {} {\bibfield
  {journal} {\bibinfo  {journal} {Physical Review D}\ }\textbf {\bibinfo
  {volume} {54}},\ \bibinfo {pages} {7163} (\bibinfo {year}
  {1996})}\BibitemShut {NoStop}%
\bibitem [{\citenamefont {Bigazzi}\ \emph {et~al.}(2021)\citenamefont
  {Bigazzi}, \citenamefont {Caddeo}, \citenamefont {Canneti},\ and\
  \citenamefont {Cotrone}}]{bigazzi2021bubble}%
  \BibitemOpen
  \bibfield  {author} {\bibinfo {author} {\bibfnamefont {F.}~\bibnamefont
  {Bigazzi}}, \bibinfo {author} {\bibfnamefont {A.}~\bibnamefont {Caddeo}},
  \bibinfo {author} {\bibfnamefont {T.}~\bibnamefont {Canneti}},\ and\ \bibinfo
  {author} {\bibfnamefont {A.~L.}\ \bibnamefont {Cotrone}},\ }\bibfield
  {title} {\bibinfo {title} {Bubble wall velocity at strong coupling},\
  }\href@noop {} {\bibfield  {journal} {\bibinfo  {journal} {Journal of High
  Energy Physics}\ }\textbf {\bibinfo {volume} {2021}},\ \bibinfo {pages} {1}
  (\bibinfo {year} {2021})}\BibitemShut {NoStop}%
\bibitem [{\citenamefont {Steinhardt}(1982)}]{PhysRevD.25.2074}%
  \BibitemOpen
  \bibfield  {author} {\bibinfo {author} {\bibfnamefont {P.~J.}\ \bibnamefont
  {Steinhardt}},\ }\bibfield  {title} {\bibinfo {title} {Relativistic
  detonation waves and bubble growth in false vacuum decay},\ }\href
  {https://doi.org/10.1103/PhysRevD.25.2074} {\bibfield  {journal} {\bibinfo
  {journal} {Phys. Rev. D}\ }\textbf {\bibinfo {volume} {25}},\ \bibinfo
  {pages} {2074} (\bibinfo {year} {1982})}\BibitemShut {NoStop}%
\bibitem [{\citenamefont {Espinosa}\ \emph {et~al.}(2010)\citenamefont
  {Espinosa}, \citenamefont {Konstandin}, \citenamefont {No},\ and\
  \citenamefont {Servant}}]{espinosa2010energy}%
  \BibitemOpen
  \bibfield  {author} {\bibinfo {author} {\bibfnamefont {J.~R.}\ \bibnamefont
  {Espinosa}}, \bibinfo {author} {\bibfnamefont {T.}~\bibnamefont
  {Konstandin}}, \bibinfo {author} {\bibfnamefont {J.~M.}\ \bibnamefont {No}},\
  and\ \bibinfo {author} {\bibfnamefont {G.}~\bibnamefont {Servant}},\
  }\bibfield  {title} {\bibinfo {title} {Energy budget of cosmological
  first-order phase transitions},\ }\href@noop {} {\bibfield  {journal}
  {\bibinfo  {journal} {Journal of Cosmology and Astroparticle Physics}\
  }\textbf {\bibinfo {volume} {2010}}\bibinfo  {number} { (06)},\ \bibinfo
  {pages} {028}}\BibitemShut {NoStop}%
\bibitem [{\citenamefont {Caprini}\ \emph {et~al.}(2016)\citenamefont
  {Caprini}, \citenamefont {Hindmarsh}, \citenamefont {Huber}, \citenamefont
  {Konstandin}, \citenamefont {Kozaczuk}, \citenamefont {Nardini},
  \citenamefont {No}, \citenamefont {Petiteau}, \citenamefont {Schwaller},\
  and\ \citenamefont {Servant}}]{RN69}%
  \BibitemOpen
\bibfield  {number} {  }\bibfield  {author} {\bibinfo {author} {\bibfnamefont
  {C.}~\bibnamefont {Caprini}}, \bibinfo {author} {\bibfnamefont
  {M.}~\bibnamefont {Hindmarsh}}, \bibinfo {author} {\bibfnamefont
  {S.}~\bibnamefont {Huber}}, \bibinfo {author} {\bibfnamefont
  {T.}~\bibnamefont {Konstandin}}, \bibinfo {author} {\bibfnamefont
  {J.}~\bibnamefont {Kozaczuk}}, \bibinfo {author} {\bibfnamefont
  {G.}~\bibnamefont {Nardini}}, \bibinfo {author} {\bibfnamefont {J.~M.}\
  \bibnamefont {No}}, \bibinfo {author} {\bibfnamefont {A.}~\bibnamefont
  {Petiteau}}, \bibinfo {author} {\bibfnamefont {P.}~\bibnamefont
  {Schwaller}},\ and\ \bibinfo {author} {\bibfnamefont {G.}~\bibnamefont
  {Servant}},\ }\bibfield  {title} {\bibinfo {title} {Science with the
  space-based interferometer elisa. ii: Gravitational waves from cosmological
  phase transitions},\ }\href@noop {} {\bibfield  {journal} {\bibinfo
  {journal} {Journal of cosmology and astroparticle physics}\ }\textbf
  {\bibinfo {volume} {2016}},\ \bibinfo {pages} {001} (\bibinfo {year}
  {2016})}\BibitemShut {NoStop}%
\bibitem [{\citenamefont {Kamionkowski}\ \emph {et~al.}(1994)\citenamefont
  {Kamionkowski}, \citenamefont {Kosowsky},\ and\ \citenamefont
  {Turner}}]{kamionkowski1994gravitational}%
  \BibitemOpen
  \bibfield  {author} {\bibinfo {author} {\bibfnamefont {M.}~\bibnamefont
  {Kamionkowski}}, \bibinfo {author} {\bibfnamefont {A.}~\bibnamefont
  {Kosowsky}},\ and\ \bibinfo {author} {\bibfnamefont {M.~S.}\ \bibnamefont
  {Turner}},\ }\bibfield  {title} {\bibinfo {title} {Gravitational radiation
  from first-order phase transitions},\ }\href@noop {} {\bibfield  {journal}
  {\bibinfo  {journal} {Physical Review D}\ }\textbf {\bibinfo {volume} {49}},\
  \bibinfo {pages} {2837} (\bibinfo {year} {1994})}\BibitemShut {NoStop}%
\bibitem [{\citenamefont {Hindmarsh}\ \emph {et~al.}(2015)\citenamefont
  {Hindmarsh}, \citenamefont {Huber}, \citenamefont {Rummukainen},\ and\
  \citenamefont {Weir}}]{RN68}%
  \BibitemOpen
  \bibfield  {author} {\bibinfo {author} {\bibfnamefont {M.}~\bibnamefont
  {Hindmarsh}}, \bibinfo {author} {\bibfnamefont {S.~J.}\ \bibnamefont
  {Huber}}, \bibinfo {author} {\bibfnamefont {K.}~\bibnamefont {Rummukainen}},\
  and\ \bibinfo {author} {\bibfnamefont {D.~J.}\ \bibnamefont {Weir}},\
  }\bibfield  {title} {\bibinfo {title} {Numerical simulations of acoustically
  generated gravitational waves at a first order phase transition},\
  }\href@noop {} {\bibfield  {journal} {\bibinfo  {journal} {Physical Review
  D}\ }\textbf {\bibinfo {volume} {92}},\ \bibinfo {pages} {123009} (\bibinfo
  {year} {2015})}\BibitemShut {NoStop}%
\bibitem [{\citenamefont {Eichhorn}\ \emph
  {et~al.}(2021{\natexlab{b}})\citenamefont {Eichhorn}, \citenamefont {Lumma},
  \citenamefont {Pawlowski}, \citenamefont {Reichert},\ and\ \citenamefont
  {Yamada}}]{RN15}%
  \BibitemOpen
  \bibfield  {author} {\bibinfo {author} {\bibfnamefont {A.}~\bibnamefont
  {Eichhorn}}, \bibinfo {author} {\bibfnamefont {J.}~\bibnamefont {Lumma}},
  \bibinfo {author} {\bibfnamefont {J.~M.}\ \bibnamefont {Pawlowski}}, \bibinfo
  {author} {\bibfnamefont {M.}~\bibnamefont {Reichert}},\ and\ \bibinfo
  {author} {\bibfnamefont {M.}~\bibnamefont {Yamada}},\ }\bibfield  {title}
  {\bibinfo {title} {Universal gravitational-wave signatures from heavy new
  physics in the electroweak sector},\ }\href
  {https://doi.org/10.1088/1475-7516/2021/05/006} {\bibfield  {journal}
  {\bibinfo  {journal} {Journal of Cosmology and Astroparticle Physics}\
  }\textbf {\bibinfo {volume} {2021}},\ \bibinfo {pages} {006}}\BibitemShut
  {NoStop}%
\bibitem [{\citenamefont {Hashino}\ \emph {et~al.}(2022)\citenamefont
  {Hashino}, \citenamefont {Kanemura},\ and\ \citenamefont
  {Takahashi}}]{Hashino:2021qoq}%
  \BibitemOpen
  \bibfield  {author} {\bibinfo {author} {\bibfnamefont {K.}~\bibnamefont
  {Hashino}}, \bibinfo {author} {\bibfnamefont {S.}~\bibnamefont {Kanemura}},\
  and\ \bibinfo {author} {\bibfnamefont {T.}~\bibnamefont {Takahashi}},\
  }\bibfield  {title} {\bibinfo {title} {{Primordial black holes as a probe of
  strongly first-order electroweak phase transition}},\ }\href
  {https://doi.org/10.1016/j.physletb.2022.137261} {\bibfield  {journal}
  {\bibinfo  {journal} {Phys. Lett. B}\ }\textbf {\bibinfo {volume} {833}},\
  \bibinfo {pages} {137261} (\bibinfo {year} {2022})},\ \Eprint
  {https://arxiv.org/abs/2111.13099} {arXiv:2111.13099 [hep-ph]} \BibitemShut
  {NoStop}%
\bibitem [{\citenamefont {Lewicki}\ \emph {et~al.}(2023)\citenamefont
  {Lewicki}, \citenamefont {Toczek},\ and\ \citenamefont
  {Vaskonen}}]{Lewicki:2023ioy}%
  \BibitemOpen
  \bibfield  {author} {\bibinfo {author} {\bibfnamefont {M.}~\bibnamefont
  {Lewicki}}, \bibinfo {author} {\bibfnamefont {P.}~\bibnamefont {Toczek}},\
  and\ \bibinfo {author} {\bibfnamefont {V.}~\bibnamefont {Vaskonen}},\
  }\bibfield  {title} {\bibinfo {title} {{Primordial black holes from strong
  first-order phase transitions}},\ }\href
  {https://doi.org/10.1007/JHEP09(2023)092} {\bibfield  {journal} {\bibinfo
  {journal} {JHEP}\ }\textbf {\bibinfo {volume} {09}},\ \bibinfo {pages}
  {092}},\ \Eprint {https://arxiv.org/abs/2305.04924} {arXiv:2305.04924
  [astro-ph.CO]} \BibitemShut {NoStop}%
\bibitem [{\citenamefont {Gouttenoire}\ and\ \citenamefont
  {Volansky}(2024)}]{Gouttenoire:2023naa}%
  \BibitemOpen
  \bibfield  {author} {\bibinfo {author} {\bibfnamefont {Y.}~\bibnamefont
  {Gouttenoire}}\ and\ \bibinfo {author} {\bibfnamefont {T.}~\bibnamefont
  {Volansky}},\ }\bibfield  {title} {\bibinfo {title} {{Primordial black holes
  from supercooled phase transitions}},\ }\href
  {https://doi.org/10.1103/PhysRevD.110.043514} {\bibfield  {journal} {\bibinfo
   {journal} {Phys. Rev. D}\ }\textbf {\bibinfo {volume} {110}},\ \bibinfo
  {pages} {043514} (\bibinfo {year} {2024})},\ \Eprint
  {https://arxiv.org/abs/2305.04942} {arXiv:2305.04942 [hep-ph]} \BibitemShut
  {NoStop}%
\bibitem [{\citenamefont {Zhang}\ \emph {et~al.}(2025)\citenamefont {Zhang},
  \citenamefont {Hashino}, \citenamefont {Ishida},\ and\ \citenamefont
  {Matsuzaki}}]{Zhang:2025kbu}%
  \BibitemOpen
  \bibfield  {author} {\bibinfo {author} {\bibfnamefont {H.-X.}\ \bibnamefont
  {Zhang}}, \bibinfo {author} {\bibfnamefont {K.}~\bibnamefont {Hashino}},
  \bibinfo {author} {\bibfnamefont {H.}~\bibnamefont {Ishida}},\ and\ \bibinfo
  {author} {\bibfnamefont {S.}~\bibnamefont {Matsuzaki}},\ }\bibfield  {title}
  {\bibinfo {title} {{Significance of soft-scale breaking on primordial black
  hole production in Coleman-Weinberg type supercooling-phase transition}},\
  }\href {https://doi.org/10.1007/JHEP10(2025)207} {\bibfield  {journal}
  {\bibinfo  {journal} {JHEP}\ }\textbf {\bibinfo {volume} {10}},\ \bibinfo
  {pages} {207}},\ \Eprint {https://arxiv.org/abs/2506.23752} {arXiv:2506.23752
  [hep-ph]} \BibitemShut {NoStop}%
\bibitem [{\citenamefont {Jiang}\ \emph {et~al.}(2025)\citenamefont {Jiang},
  \citenamefont {Guan}, \citenamefont {Kawaguchi}, \citenamefont {Matsuzaki},
  \citenamefont {Tomiya},\ and\ \citenamefont {Zhang}}]{Jiang:2025ofd}%
  \BibitemOpen
  \bibfield  {author} {\bibinfo {author} {\bibfnamefont {Z.-l.}\ \bibnamefont
  {Jiang}}, \bibinfo {author} {\bibfnamefont {Y.}~\bibnamefont {Guan}},
  \bibinfo {author} {\bibfnamefont {M.}~\bibnamefont {Kawaguchi}}, \bibinfo
  {author} {\bibfnamefont {S.}~\bibnamefont {Matsuzaki}}, \bibinfo {author}
  {\bibfnamefont {A.}~\bibnamefont {Tomiya}},\ and\ \bibinfo {author}
  {\bibfnamefont {H.-X.}\ \bibnamefont {Zhang}},\ }\bibfield  {title} {\bibinfo
  {title} {{Axionlike particle-assisted supercooling chiral phase transition in
  QCD: Identifying Coleman-Weinberg type-chiral phase transition in QCD-like
  scenarios}},\ }\href@noop {} {\  (\bibinfo {year} {2025})},\ \Eprint
  {https://arxiv.org/abs/2507.18016} {arXiv:2507.18016 [hep-ph]} \BibitemShut
  {NoStop}%
\end{thebibliography}%

\end{document}